\documentclass[twocolumn,floatfix,superscriptaddress,aps,prx]{revtex4-1}
\usepackage{graphicx}
\usepackage{amsmath,mathtools}
\usepackage{amssymb}
\usepackage{bm}
\usepackage{hyperref}
\usepackage{color}

\begin{document}

\title{Quantum tunneling dynamics in a complex-valued Sachdev-Ye-Kitaev model quench-coupled to a cool bath}

\begin{abstract}
The Sachdev-Ye-Kitaev (SYK) model describes interacting fermionic zero modes in zero spatial dimensions, e.g. quantum dot, with interactions strong enough to completely washout quasiparticle excitations in the infrared.
In this paper, we consider the complex-valued SYK model at initial temperature $T$ and chemical potential $\mu$ coupled to a large reservoir by a quench at time $t=0$. The reservoir is kept at zero temperature and charge neutrality. We find that the dynamics of the discharging process of the SYK quantum dot reveals a distinctive characteristic of the SYK non-Fermi liquid (nFl) state. In particular, we focus on the tunneling current induced by the quench. We show that the temperature dependent contribution to the current's half-life scales linearly in $T$ at low temperatures for the SYK nFl state, while for the Fermi liquid it scales as $T^2$.
\end{abstract}

\author{Y. Cheipesh}
\affiliation{Instituut-Lorentz, Universiteit Leiden, P.O. Box 9506, 2300 RA Leiden, The Netherlands}

\author{A. I. Pavlov}
\affiliation{The Abdus Salam International Centre for Theoretical Physics (ICTP) Strada Costiera 11, I-34151 Trieste, Italy}

\author{V. Ohanesjan}
\affiliation{Instituut-Lorentz, Universiteit Leiden, P.O. Box 9506, 2300 RA Leiden, The Netherlands}

\author{K. Schalm}
\affiliation{Instituut-Lorentz, Universiteit Leiden, P.O. Box 9506, 2300 RA Leiden, The Netherlands}

\author{N. V. Gnezdilov}
\email{n.gnezdilov@ufl.edu}
\affiliation{Department of Physics, University of Florida, Gainesville, FL 32611, USA}
\affiliation{Department of Physics, Yale University, New Haven, CT 06520, USA}

\date{\today}

\maketitle

\section{Introduction}\label{sec:intro}

Non-equilibrium dynamics of the celebrated Sachdev-Ye-Kitaev (SYK) model \cite{Kitaev2015Simple, Sachdev1993Gapless}  -- dual to a black hole in a two-dimensional anti-de Sitter space -- instantaneously coupled to a larger cold media has been recently scrutinized \cite{Almheiri2019Universal, Zhang2019Evaporation} intending to mimic black hole evaporation \cite{Hawking1974Black, Hawking1976Breakdown, Page1993Information, Almheiri2019Entropy, Almheiri2019Islands} in a compact quantum mechanical setup. Alongside, several platforms have been proposed for experimental realization of the SYK model: as a low-energy effective description of a topological insulator/superconductor interface with an irregular opening \cite{Pikulin2017Black}, Majorana wires coupled through a disordered quantum dot \cite{Chew2017Approximating}, ultracold atoms trapped in optical lattices \cite{Danshita2016Creating,Wei2021Optical}, graphene flake with a random boundary \cite{Chen2018Quantum}, and digital quantum simulation \cite{Garcia2017Digital, Luo2019Quantum, Babbush2019Quantum}.
In this context, opening up the system to an outer environment arises naturally as the ``black-hole chip'' \cite{Pikulin2017Black} is necessarily in contact with a substrate and probes. 

Once the system is opened due to quench-coupling, it starts to equilibrate with the external reservoir. Of particular interest is how the initial shock and the subsequent equilibration affects the initial SYK state and transport observables. 
The SYK model describes strongly interacting fermions in $(0+1)$-dimensions. As such, it can be considered as a quantum dot that is usually characterized via tunneling current. In this manuscript, we consider the complex SYK model \cite{Sachdev2015Bekenstein, Gu2019Notes} abruptly coupled to a zero temperature bath. We input the initial electrochemical potential in the SYK subsystem to enable quantum charge tunneling apart from the temperature drop between the SYK dot and the reservoir \cite{Almheiri2019Universal, Zhang2019Evaporation}. 
Unlike equilibrium transport in the SYK quantum dot coupled to metallic leads \cite{Gnezdilov2018Low, Can2019Charge, Altland2019Sachdev, Kruchkov2020Thermoelectric, Pavlov2020Quantum}, we are focused on the time evolution of both spectral properties and the tunneling current.

It was indicated earlier that right after the quench the SYK subsystem surprisingly heats up despite coupling to the colder bath \cite{Almheiri2019Universal, Zhang2019Evaporation} and cools down later equilibrating with the reservoir's temperature. In the holographic picture this initial heating is aligned with the increase of the subsystem energy that accompanies the information carried by the quench-induced shock-wave falling into the black hole \cite{Almheiri2019Entropy}. We recover this result in the absence of a potential difference and confirm that the applied quench protocol cools down the SYK dot preserving an exotic SYK non-Fermi liquid phase after the relaxation. 
Proceeding to transport, we analyze the tunneling current evolution at low temperatures. We observe numerically that the current half-life -- the time required for current to relax back to half its maximum value -- growths linearly with the initial temperature of the SYK quantum dot. 
In contrast, replacing the SYK subsystem with a disordered Fermi liquid leads to a quadratic temperature increment of the current's half-life. This enables one to distinguish the SYK non-Fermi liquid from a more common disordered phase by means of the quench-tunneling protocol.

\section{The model}\label{sec:model}

We begin our analysis with the SYK model in thermal equilibrium (chemical potential $\mu$, temperature $T$) coupled to a reservoir at zero chemical potential and zero temperature via tunneling term at time $t=0$. The Hamiltonian reads 
\begin{align}
H&=H_{SY\!K}+H_{res}+ \theta(t) H_{tun}, \label{H_full}
\end{align}
\begin{align}
H_{SY\!K}&=\frac{1}{\left(2N\right)^{3/2}}\!\sum_{i,j,k,l=1}^N \! J_{ij;kl} c_i^\dag c_j^\dag c_k c_l - \mu \sum_{i=1}^N c_i^\dag c_i, \label{SYK} 
\end{align}
\begin{align}
H_{res}&=\frac{1}{\sqrt{M}}\!\sum_{\alpha,\beta=1}^M\!  \xi_{\alpha\beta}  \psi_\alpha^\dag \psi_\beta + h.c., \label{H_bath}  
\end{align}
\begin{align}
H_{tun}&=\frac{1}{\left(N M\right)^{1/4}}\!\sum_{i=1}^N\!\sum_{\alpha=1}^M\!  \lambda_{i\alpha}  c_i^\dag \psi_\alpha + h.c. , \label{H_tun}
\end{align}
where $J_{ij;kl}=J^*_{kl;ij}=-J_{ji;kl}=-J_{ij;lk}$, $\xi_{\alpha\beta}$, and $\lambda_{i\alpha}$ are Gaussian random variables with finite variances $\overline{|J_{ij;kl}|^2}=J^2$, $\overline{|\xi_{\alpha\beta}|^2}=\xi^2$, $\overline{|\lambda_{i\alpha}|^2}=\lambda^2$ and zero means. Below we assume the reservoir much larger than the SYK subsystem, which imposes $M \gg N$ for the modes numbers. The charging energy \cite{Altland2019Sachdev, Pavlov2020Quantum, Khveshchenko2020SET, Khveshchenko2020Connecting} is supposed to be negligible comparing to the SYK band-width $J$.  

The conventional way to address non-equilibrium dynamics of a quantum many-body system is solving Kadanoff-Baym (KB) equations for the two-point functions $G^\gtrless (t,t')=-i N^{-1} \sum\limits_{i=1}^N \langle c_i(t_\mp) \bar{c}_i(t'_\pm) \rangle$, where $\pm$ denotes the top/bottom branches of the Keldysh time contour \cite{Kamenev2005Course}. Inasmuch as Schwinger-Keldysh formalism has been widely applied to the SYK model in both thermalization \cite{Eberlein2017Quantum, Bhattacharya2019Quantum, Zhang2019Evaporation, Almheiri2019Universal, Kuhlenkamp2020Periodically, Haldar2020Quench} and transport \cite{Son2017Strongly, Gnezdilov2018Low, Can2019Charge} context, we leave the detailed derivation for Appendix \ref{app:KB} and proceed straight to the Kadanoff-Baym equations that hold in the large $N, M$ limit:
\begin{align} \nonumber
\left(i \partial_t+\mu\right) G^\gtrless (t,t') =&  \int_{-\infty}^{+\infty}\!\!du \Big( \Sigma_R(t,u) G^\gtrless (u,t') \\&+ \Sigma^\gtrless (t,u) G_A(u,t')\Big), \label{eqKBt}
\end{align}
\begin{align} \nonumber
\left(-i \partial_{t'}+\mu\right) G^\gtrless (t,t')  = & \int_{-\infty}^{+\infty}\!\!du \Big( G_R(t,u) \Sigma^\gtrless (u,t') \\ &+ G^\gtrless (t,u) \Sigma_A(u,t')\Big), \label{eqKBtp}
\end{align}
The self-energy 
\begin{align} \nonumber
\Sigma^\gtrless (t,t')  =& J^2 G^\gtrless(t,t')^2 G^\lessgtr(t'\!,t) \\ &+ \sqrt{p} \, \lambda^2 \theta(t) \theta(t') Q^\gtrless(t,t') \label{Sigma}
\end{align}
includes the contribution of the cool-bath as a time dependent background 
\begin{align}
Q^\gtrless(t,t')&=-\frac{\mathbf{H}_1\!\left(2\xi (t-t')\right) \pm i J_1\!\left(2\xi (t-t')\right)}{2\xi(t-t')} \label{Qgl}
\end{align}
expressed through Struve $\mathbf{H}_1$ and Bessel $J_1$ functions \cite{Abramowitz1964Handbook}; see Appendix \ref{app:KB}. Here we introduce the  ratio $p=M/N$ and limit ourselves to the large reservoir case $p \gg 1$. Below we assume $\xi = J$ for brevity. 

The initial state of the system is settled by the thermal state of the bare SYK model (\ref{SYK}) in absence of coupling to the reservoir.
At the moment of quench the SYK subsystem (\ref{SYK}) begins to deviate from the initial thermal state until it finally thermalizes at late times. Characterizing thermalization dynamics requires notion of the retarded, advanced, and Keldysh Green's functions 
\begin{align}
G_R(t,t') & = \theta(t-t') \Big(G^>(t,t') - G^<(t,t')\Big), \label{GRgl}\\
G_A(t,t') & = -\theta(t'\!-t)  \Big(G^>(t,t') - G^<(t,t')\Big), \label{GAgl}\\
G_K(t,t') & = G^>(t,t') + G^<(t,t') \label{GKgl}
\end{align}
expressed above in terms of the ``greater'' and ``lesser'' components.  The same rules (\ref{GRgl}--\ref{GKgl}) apply to the self-energy (\ref{Sigma}). 

The Green's functions are found numerically from the KB equations (\ref{eqKBt},\ref{eqKBtp}) with the self-energies (\ref{Sigma},\ref{Qgl}). At first, we calculate the equilibrium Green's functions of the bare SYK model using an iterative approach \cite{Malitsky2019Golden, Cheipesh2019Reentrant}. We apply an extra constraint manifesting the fluctuation-dissipation relation at initial temperature and chemical potential \footnote{In thermal equilibrium the fluctuation dissipation relation states \cite{Kamenev2005Course}: $G_K(\omega) = 2 i \, \mathrm{Im}G_R(\omega) \tanh\dfrac{\omega-\mu}{2T}$}. The equilibrium Green's functions set the initial condition for the Kadanoff-Baym equations and evolve as follows: the integrals in the KB equations are computed with the trapezoidal rule and the remaining differential equations are solved by the predictor-corrector scheme.  The corrector adjusts self-consistently at every iteration \cite{Eberlein2017Quantum, Bhattacharya2019Quantum}. For the spectral properties we use the two-dimensional time grid with a step $\delta t = 0.02$ and $n \sim 10^4$ points in each direction, while for the transport calculations the numerical grid is more refined $\delta t = 0.005$ but has a smaller size $n \sim 10^3$.

\section{Relaxation after the quench}\label{sec:relaxation}

\begin{figure*}[t!!]
\center
\includegraphics[width=0.464\linewidth]{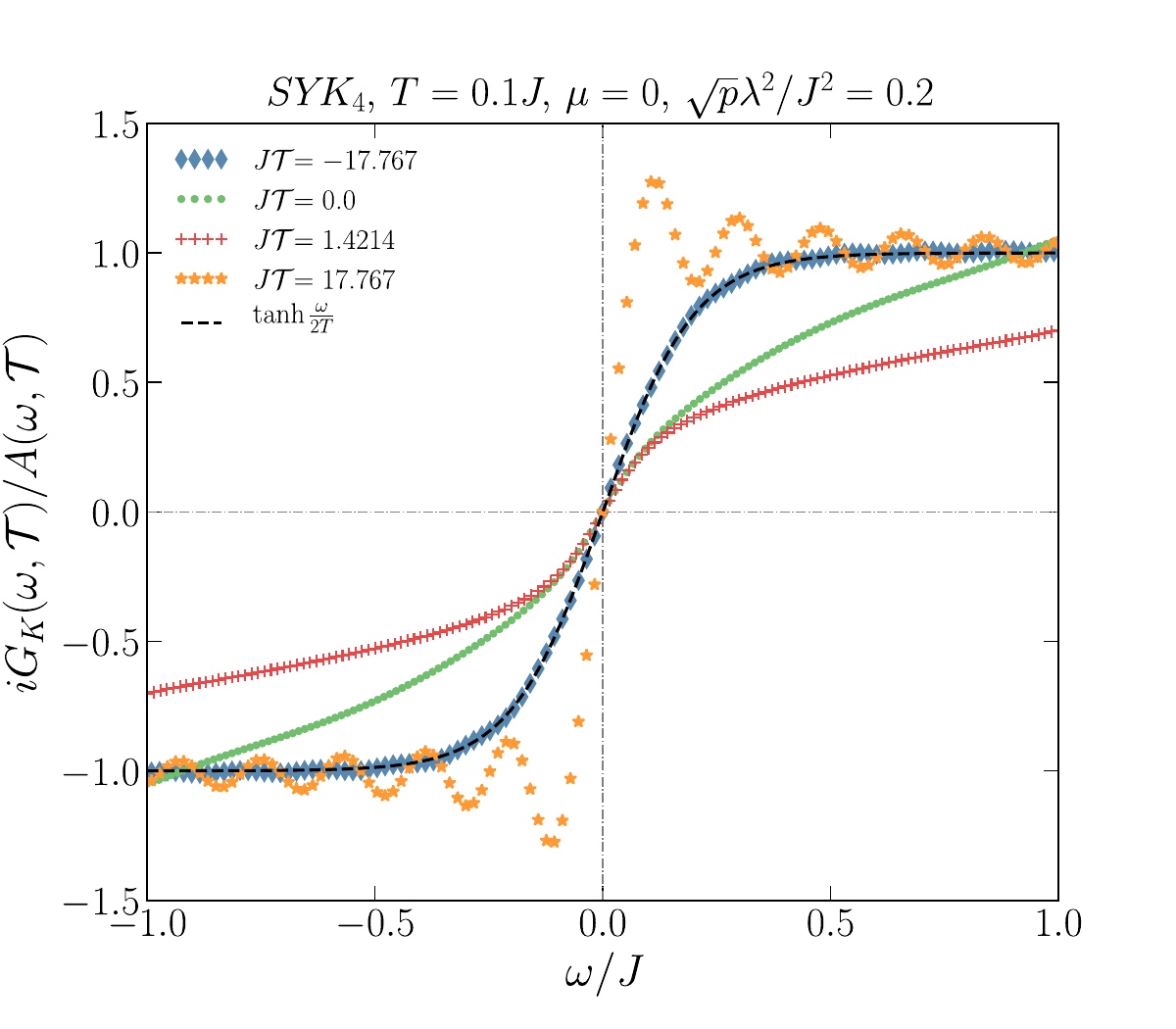} \quad
\includegraphics[width=0.464\linewidth]{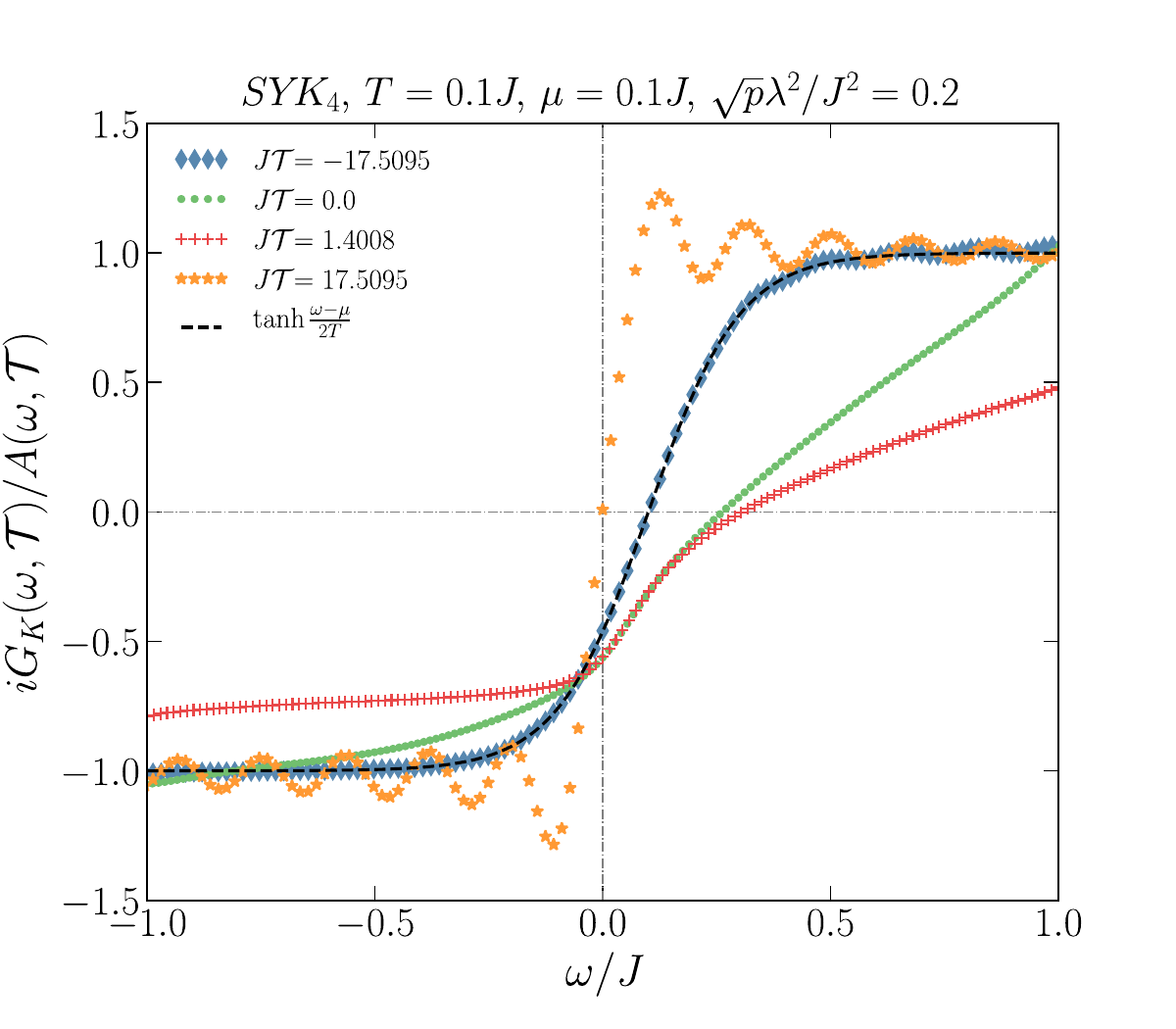}
\includegraphics[width=0.464\linewidth]{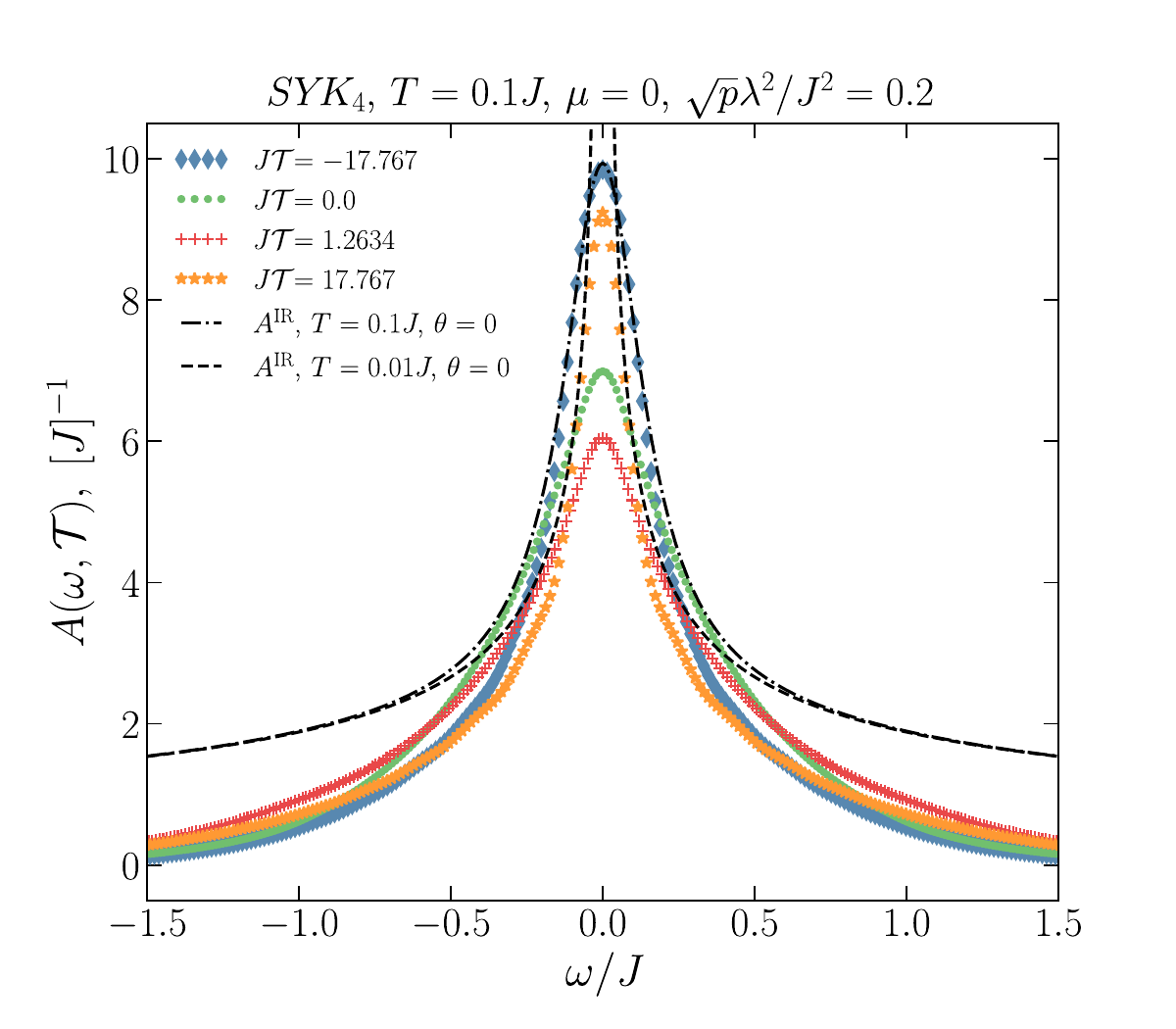} \quad
\includegraphics[width=0.464\linewidth]{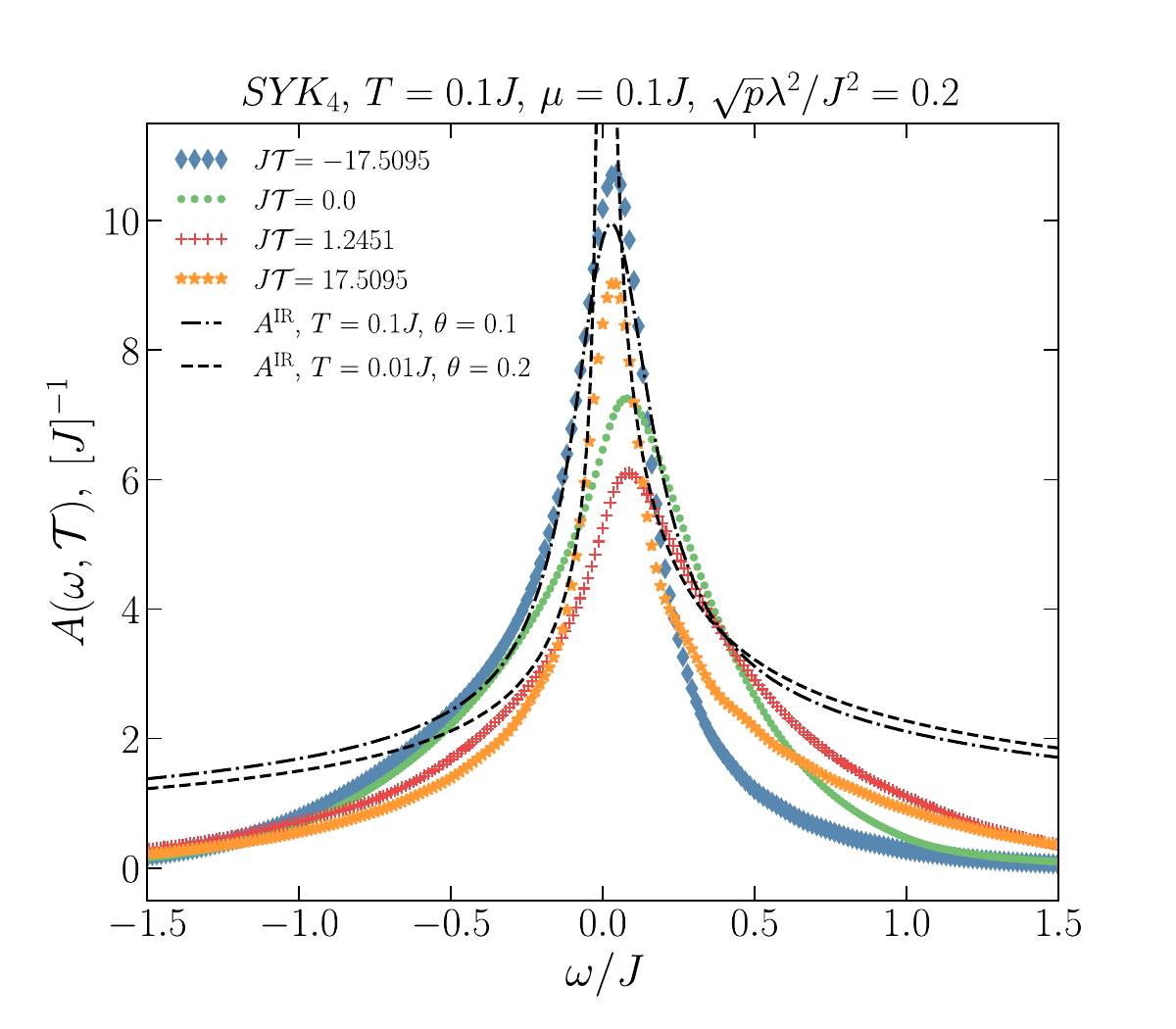}
\caption{\small \label{fig:thermal}  \textbf{[Top] Deviation of the SYK subsystem from the initial thermal state:} ratio between the Keldysh Green's function and the spectral function of the SYK model at charge neutrality (\textit{Left panel}) and at finite chemical potential (\textit{Right panel}). The equilibrium distribution functions at the initial temperature are profiled with the dashed lines. The oscillations noticeable in the orange curves have a numerical origin, viz. the quality of the computation depends on the size and refinement of the time grid.  The time grid is designated in the $t,t'$ space, while Fourier transform is done along diagonal $\tau = t-t'$. Ergo, the $\tau$-lattices differ by length for separate slices of $\mathcal{T}$. Extension and refinement of the time grid suppress the oscillations.
\textbf{[Bottom] Spectral function of the SYK model} as a function of frequency at charge neutrality (\textit{Left panel}) and at finite chemical potential (\textit{Right panel}). The dashed/dash-dot lines show the equilibrium SYK spectral function in the infrared regime for different parameters.}
\end{figure*}

In a while after the quench the system relaxes and approaches a thermal state.
To demonstrate that, we rotate the time frame $t,t'$ in the numerically computed Green's functions towards $\tau=t-t'$, $\mathcal{T}=(t+t')/2$ and make a Fourier transform along $\tau$. Indeed, the system returns to a nearly-thermal state if the extended fluctuation dissipation relation  
\begin{align}
    \frac{i G_K(\omega,\mathcal{T})}{A(\omega, \mathcal{T})} = \tanh\frac{\omega-\widetilde{\mu}(\mathcal{T})}{2\widetilde{T}(\mathcal{T})} \label{FDT}
\end{align}
is fulfilled at frequencies in the vicinity of $\widetilde{\mu}$, where $A(\omega, \mathcal{T})=-2 {\rm Im} G_R(\omega, \mathcal{T})$ is the SYK spectral function. In contrast to the equilibrium case, the extended fluctuation dissipation relation (\ref{FDT}) is manifestly time dependent via the ``centre of mass'' coordinate $\mathcal{T}$ which enters the effective temperature $\widetilde{T}$ and chemical potential $\widetilde{\mu}$. Overall, the ratio (\ref{FDT}) determines the effective distribution function of the SYK fermions in a quasi-equilibrium state, since $\tanh\frac{\omega-\widetilde{\mu}}{2\widetilde{T}}= 1 - 2 n_{\rm F}(\omega -\widetilde{\mu}, T)$, where $n_{\rm F}$ is the Fermi distribution function.  

\begin{figure}[t!!]
\center
\includegraphics[width=0.928\linewidth]{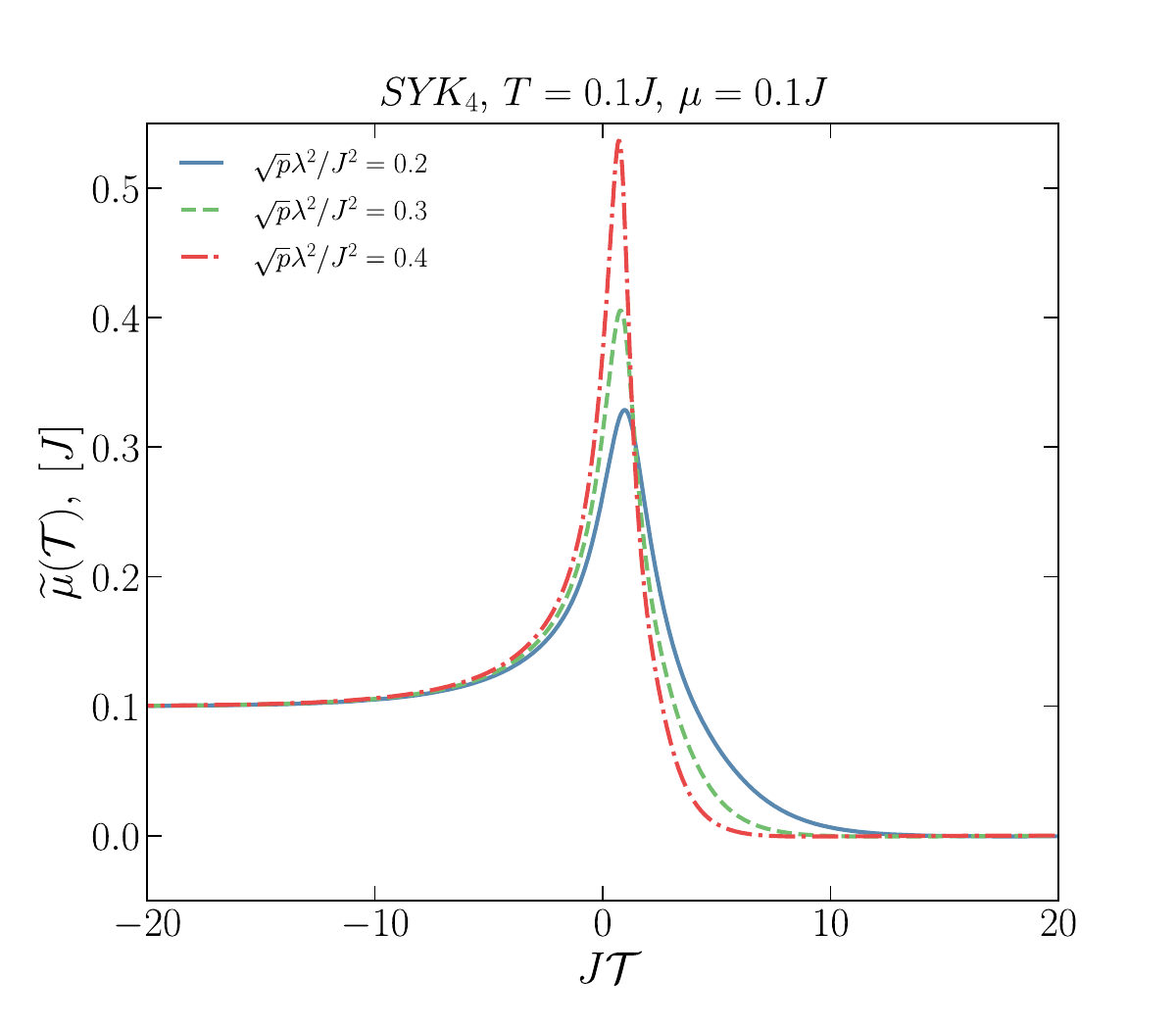} 
\caption{\small \label{fig:mu}  \textbf{Effective chemical potential in the SYK quantum dot} coupled to a large reservoir with $T_{res}=0$ and $\mu_{res}=0$.}
\end{figure}

The effective temperature can be extracted  from the fluctuation dissipation relation (\ref{FDT}) by an inverse slope of the Green's functions ratio 
\begin{align}
   \widetilde{T}(\mathcal{T}) =  \left(\left. \frac{\partial}{\partial \omega}  \frac{2 i G_K(\omega,\mathcal{T})}{A(\omega, \mathcal{T})} \right|_{\omega=\widetilde{\mu}}\right)^{-1}
\end{align}
at $\omega=\widetilde{\mu}$. 
Following the top panel of Fig. \ref{fig:thermal}, that shows the ratio (\ref{FDT}), one notices the temperature increase around $\mathcal{T}=0$, in spite of coupling to a colder reservoir. The initial temperature increment is followed by the subsequent temperature decay to the reservoir's temperature $T=0$. This behavior was revealed earlier for the SYK model with Majorana zero-modes \cite{Almheiri2019Universal, Zhang2019Evaporation}. At late times $J\mathcal{T} \simeq 17.8$, the system clearly relaxes after the quench since the ratio (\ref{FDT}) corresponds to the Fermi distribution at low temperature. 

In comparison to the previous studies \cite{Zhang2019Evaporation, Almheiri2019Universal}, the new ingredient here is a charge imbalance between the SYK quantum dot and the cool-bath.
Thereby, we track the electrochemical potential in the SYK subsystem which changes substantially once the quench is on. The effective chemical potential $\widetilde{\mu}(\mathcal{T})$ is set by the frequency where the ratio (\ref{FDT}) turns to zero, as shown in Fig. \ref{fig:thermal} (\textit{top right}). We plot the SYK chemical potential in Fig. \ref{fig:mu}, where $\widetilde{\mu}$ originates from the initial value $\mu = 0.1 J$ in the SYK quantum dot for $\mathcal{T} \to -\infty$ and adjusts to the reservoir's $\mu_{res} = 0$ at late times $\mathcal{T} \to +\infty$. As noted in Fig. \ref{fig:mu}, the chemical potential responds to the quench with a non-monotonic behavior as a function of time $\mathcal{T}$, akin to the temperature. Note  that  the  ``centre  of  mass''  time  coordinate $\mathcal{T}$ and the actual time are not equivalent unless in a long-time limit. This explains why the chemical potential can already rise at small negative ${\cal T}$.

\begin{figure}[t!!]
\center
\includegraphics[width=0.928\linewidth]{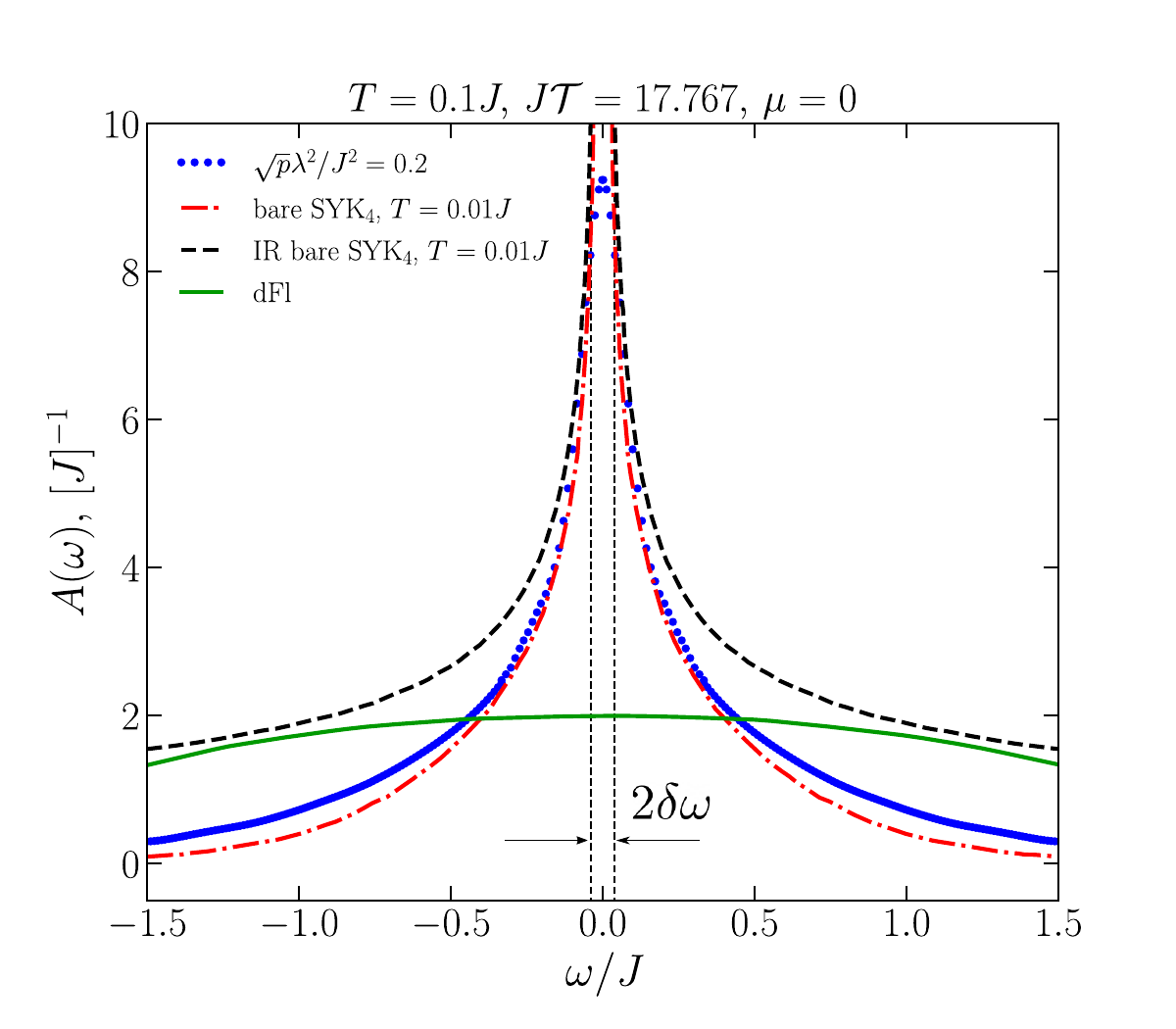} 
\caption{\small \label{fig:A02}  \textbf{Spectral of the SYK quantum dot} after the quench as a function of frequency. The blue dots show the result of the saddle-point numerics done for the evolution of the SYK subsystem with the initial temperature $T=0.1 J$ connected to a zero temperature reservoir with a coupling strength $\sqrt{p}\lambda^2/J^2 = 0.2$. The red dash-dot curve is the equilibrium saddle-point numerics for the bare SYK model at low temperature, the black dashed line is the infrared (IR) solution of the bare SYK model (\ref{GR_IR}), and the green line is the spectral function of the disordered Fermi liquid (dFl). The energy scale $\delta \omega = p \lambda^4/J^3$ indicates the region where the SYK nFl crosses over to a Fermi liquid.}
\end{figure}
 
Since the tunneling between the SYK quantum dot and the reservoir turns on not adiabatically, of importance is whether the SYK non-Fermi liquid phase survives the quench. 
We compare the SYK spectral function $A(\omega, \mathcal{T})$ a while after the quench to the equilibrium spectral function of the bare SYK model $A^{\rm I R}(\omega)= -2 {\rm Im} G^{\rm I R}_R(\omega)$ in the infrared regime $J/N \ll \omega, T \ll J$, where 
\begin{align}
    G^{\rm I R}_R(\omega) &= - i \frac{C(\theta) e^{-i \theta}}{\sqrt{2\pi J T}} \frac{\Gamma\left(\frac{1}{4}-i\frac{\omega}{2\pi T}+i\mathcal{E}\right)}{\Gamma\left(\frac{3}{4}-i\frac{\omega}{2\pi T}+i\mathcal{E}\right)} \label{GR_IR}, \\
    e^{2\pi \mathcal{E}} &=  \frac{\sin\left(\frac{\pi}{4} + \theta \right)}{\sin\left(\frac{\pi}{4}-\theta\right)}, \quad C(\theta) =\left(\frac{\pi}{\cos 2\theta} \right)^{1/4}.
\end{align}
The low-frequency asymptotic (\ref{GR_IR}), known as the conformal Green's function of the SYK model, does not explicitly depend on chemical potential. Instead, it depends on the independent parameter -- the spectral asymmetry angle \cite{Sachdev2015Bekenstein, Gu2019Notes}. The asymmetry angle $\theta$ \cite{Parcollet1998Overscreened} is nonzero away from charge neutrality ($\mu \neq 0$) and related to the charge per site on the SYK quantum dot
\begin{align}
 \langle\mathcal{Q}\rangle  = \frac{1}{N}\sum_{i=1}^N   \langle c_i^\dag c_i \rangle - \frac{1}{2} = - \frac{\theta}{\pi} - \frac{\sin 2\theta}{4}, \label{Q_av}
\end{align}
where $\langle\mathcal{Q}\rangle \in (-1/2, 1/2)$ and $\theta \in (-\pi/4, \pi/4)$ \cite{Sachdev2015Bekenstein, Gu2019Notes}.

As mentioned earlier, the system relaxes to the low-temperature Fermi distribution at $J \mathcal{T}\simeq 17.8$ (see Fig. \ref{fig:thermal} (\textit{top left})). In Fig. \ref{fig:A02} we plot the spectral function of the SYK quantum dot in this regime. The spectral function after the quench is well aligned with the bare SYK spectral function at low temperature. The SYK nFl state  is known to break down in presence of a Fermi liquid \cite{Lunkin2018Sachdev,Can2019Solvable}. Here we can estimate the timescale of the crossover to a Fermi liquid from the self-energy (\ref{Sigma}) comparing the SYK nFl and the reservoir's contributions. Indeed, substitution of the Green's functions $G(t) \propto 1/\sqrt{Jt}$ and $Q(t) \propto 1/(Jt)$ to the self-energy (\ref{Sigma}) shows that the crossover to a Fermi liquid happens for $t_{FL} \gtrsim 1/\delta\omega$, where $\delta\omega = p \lambda^4/J^3$. This implies that after relaxation from the quench the SYK nFl behavior can be read out from the spectral function for
\begin{align}
    \delta \omega \lesssim \omega \ll J. \label{nFL_region}
\end{align}
The lower bound in inequality (\ref{nFL_region}) can be suppressed as $\sqrt{p}\lambda^2/J \ll J$ in the weak tunneling limit. This observation agrees with the long timescale of the SYK nFl/Fermi liquid crossover found earlier in equilibrium studies \cite{Son2017Strongly,Chowdhury2018Translationally,Lunkin2018Sachdev,Altland2019Sachdev}.  

In Figs. \ref{fig:thermal} (\textit{top right}), \ref{fig:mu} we demonstrate that the system at finite initial $\mu$ tends to zero chemical potential in the long time limit. This is aligned with the discharging of the SYK quantum dot coupled to the large reservoir, which is kept at charge neutrality. At the level of the the equilibrium SYK Green's function (\ref{GR_IR}), this naively implies $\theta \approx 0$. However, the spectral function in Fig. \ref{fig:thermal} (\textit{bottom right}) at long times is close enough to the conformal one with non-zero asymmetry angle $\theta$. We plot the conformal spectral function with $\theta = 0.2$ as a reference. 
The origin of this mismatch may be that the asymmetry parameter $\theta$ is usually related to $\partial \mu/\partial T$  but not to the equilibrium value of the chemical potential \cite{Sachdev2015Bekenstein}. In its  turn, the temperature-independent part of the chemical potential in the SYK model is not a monotonic function of the asymmetry parameter \cite{Tikhanovskaya2020Excitation}. 
Additionally, the SYK subsystem after the quench suffers the particle leak, that may require to account not only for a self-energy shift by the real-valued $\mu$ \cite{Sachdev2015Bekenstein}, but also an extra imaginary contribution to the self-energy. This issue could lead to the renormalization of $\theta$ in the final state, which is beyond the scope of this paper.

\section{Tunneling current}\label{sec:current}

Having discussed the SYK subsystem inner properties we proceed to transport. Specifically, we focus on the tunneling current: 
\begin{align}  
\dot{\mathcal{Q}} \!=\!i [H, \mathcal{Q}] \!=\! -\frac{i}{N} \frac{\theta(t)}{(N M)^{1/4}}\!\sum_{i=1}^N\!\sum_{\alpha=1}^M\!   \lambda_{i\alpha} c_i^\dag \psi_\alpha + h.c. \label{dQdt}
\end{align}
The current's expectation value in the SYK quantum dot/cool-bath system is found from the generating functional $\ln \!Z[\chi]$ \cite{Gnezdilov2018Low}
\begin{align}
\mathcal{I} &=  \frac{1}{t_{\rm m}} \! \int_0^{t_{\rm m}} \!\! dt \langle\dot{\mathcal{Q}}(t)\rangle  = \frac{1}{t_{\rm m}} \! \left. \frac{\partial   }{\partial(i \chi)} \ln \!Z[\chi] \right|_{\chi=0} \label{I}, \\
Z[\chi] & = \left\langle\mathrm{T}_C e^{ -i \!\int_C \!dt H(\chi)} \right\rangle =  \int\!\mathcal{D}[\bar{c},c]\mathcal{D}[\bar{\psi},\psi]e^{i S[\chi]}, \label{Z_chi}
\end{align}
where $\mathrm{T}_C$ is the time ordering along the Keldysh contour, $t_{\rm m}$ is the measurement time, and $S[\chi]$ is the effective action of the model with a counting field $\chi$ \cite{Levitov1993Charge, Levitov1996Electron}. 
The counting field $\chi$ transforms the tunneling Hamiltonian 
\begin{align}
    H(\chi)&=H_{SY\!K}+H_{res}+ \theta(t) H_{tun}(\chi), \\
    H_{tun}(\chi)&=\frac{1}{\left(N M\right)^{1/4}}\!\sum_{i=1}^N\!\sum_{\alpha=1}^M\!  \lambda_{i\alpha} e^\frac{i \chi(t)}{2N}  c_i^\dag \psi_\alpha + h.c., \label{H_tun_chi}
\end{align}
so that
\begin{align}
\chi(t)=\begin{cases} \chi \quad \text{for} \quad 0<t<t_{\rm m} \\ 0 \quad \text{otherwise} \end{cases},
\end{align}
The factor of two in the coupling phase in the tunneling term (\ref{H_tun_chi}) accounts for the doubling due to the forward and backward branches of the Keldysh time contour.

\begin{figure}[t]
\center
\includegraphics[width=0.928\linewidth]{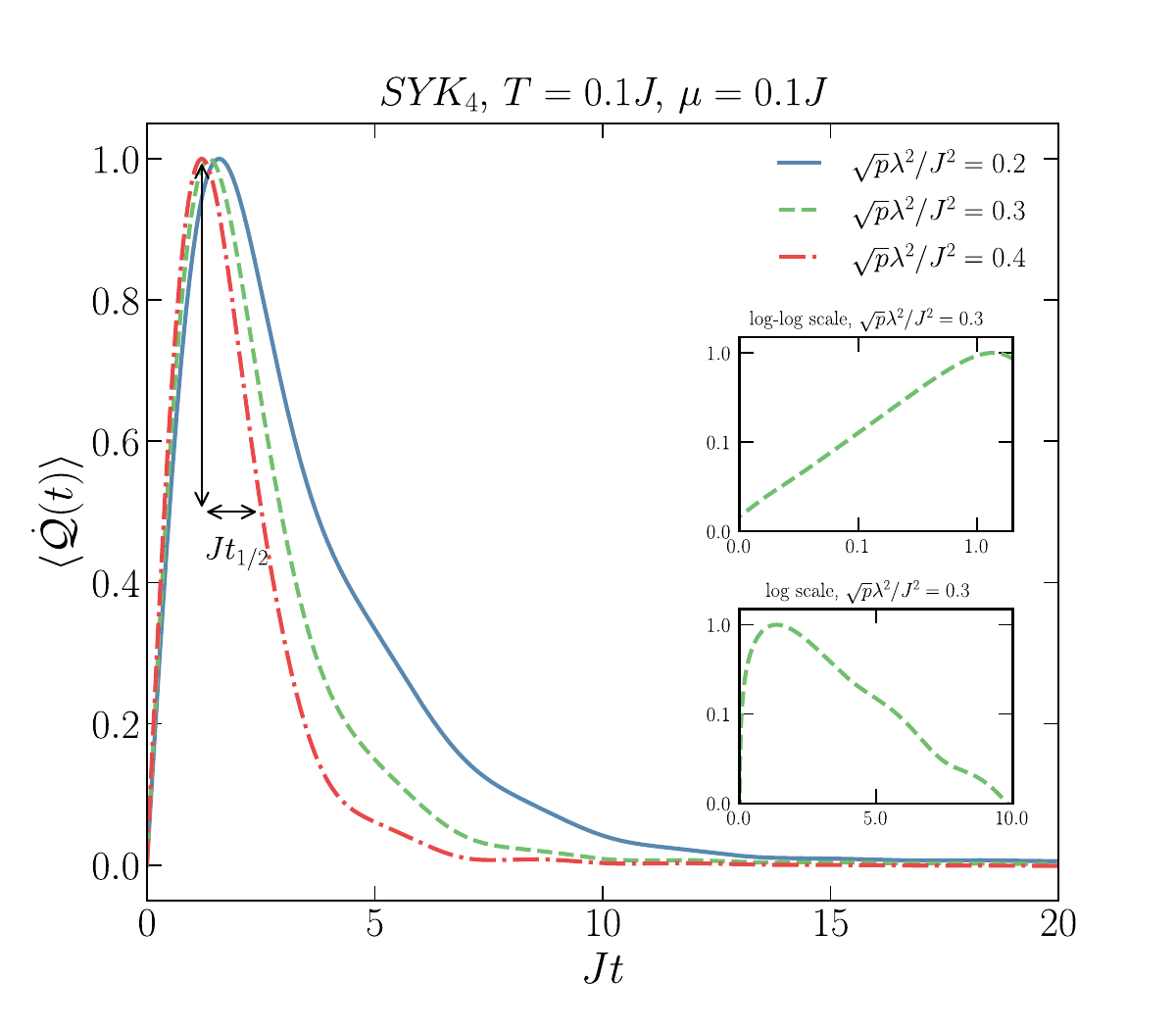} 
\caption{\small \label{fig:dQdt}  \textbf{Tunneling current} as a function of time normalized on its maximum value. The insets show time dependence of the current in log-log and log scales for $\sqrt{p}\lambda^2/J^2 =0.3$. The log-log plot reveals the initial power law increase of the tunneling current, while the log plot is consistent with the exponential decay. We illustrate the current's half-life $t_{1/2}$ for $\sqrt{p}\lambda^2/J^2 =0.4$.}
\end{figure} 

\begin{figure*}[t]
\center
\includegraphics[width=0.464\linewidth]{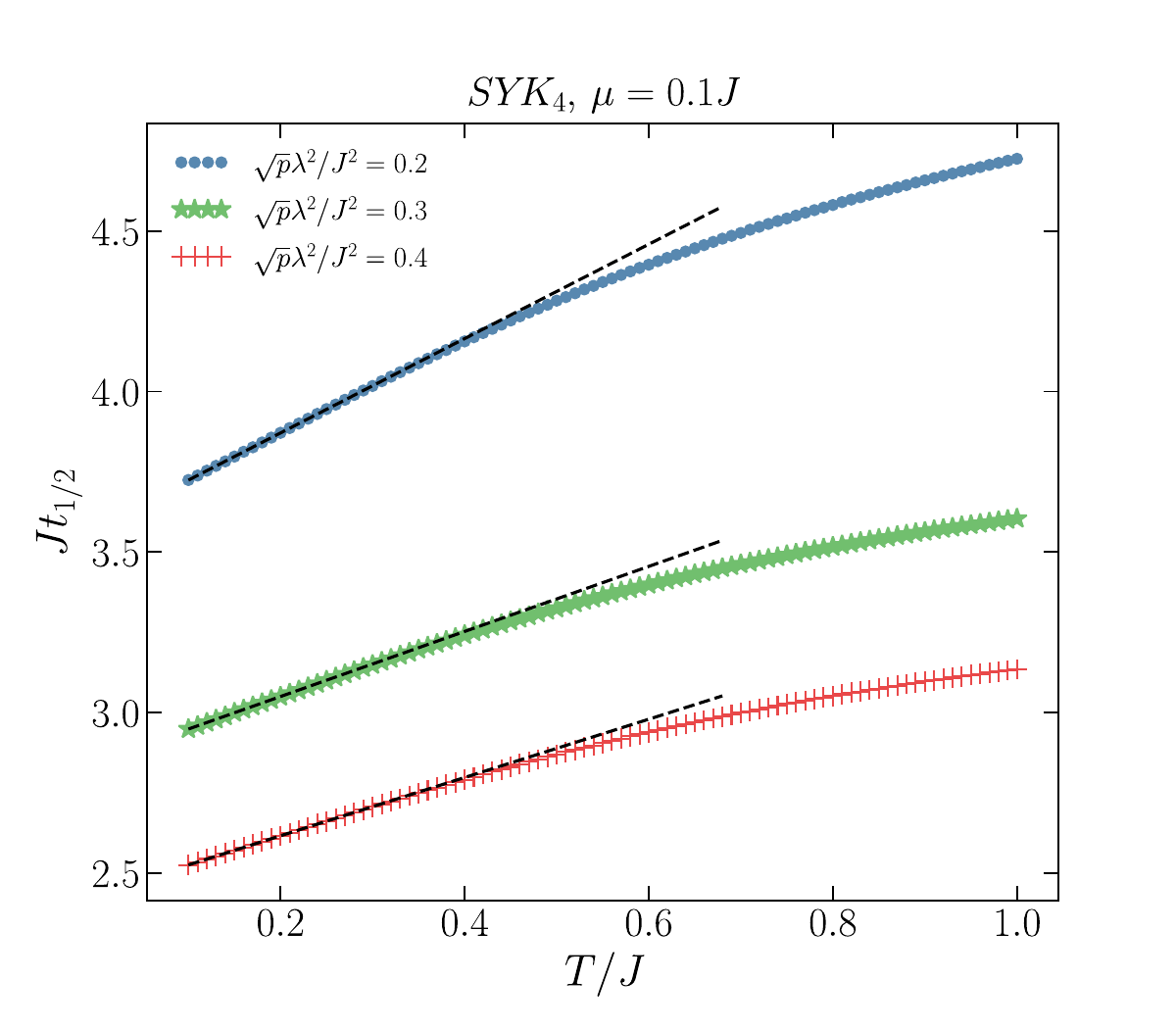} \quad
\includegraphics[width=0.464\linewidth]{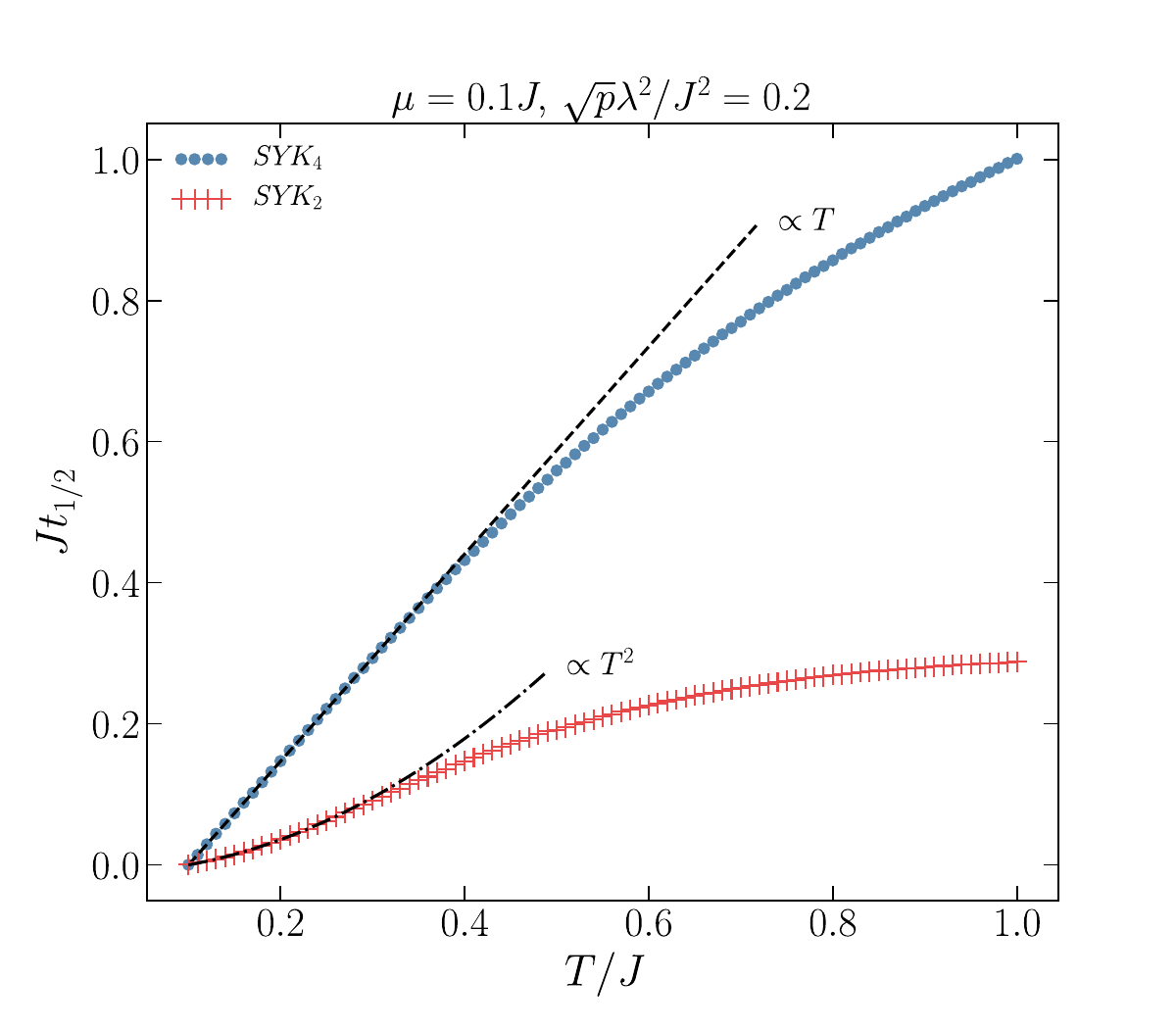}
\caption{\small \label{fig:t05}  \textbf{Half-life of the tunneling current} as a function of the initial temperature. In the \textit{left panel}, we compare the half-lives for the SYK model connected to a cool-bath for different coupling strengths. Meanwhile in the \textit{right panel}, we show the difference between \textbf{SYK$_4$} (SYK non-Fermi liquid initial state) and \textbf{SYK$_2$}  (disordered Fermi liquid initial state) behavior as a tested subsystem; the curves are shifted to the same origin for better visual comparison. The initial temperature changes from $T = 0.1 J$ to $T = J$ with a step $\delta T = 0.01 J$. The dashed/dashdot lines stand for the linear/quadratic fits made for the temperature interval $T \in [0.1 J, 0.2 J]$.}
\end{figure*}

One notices that the Hamiltonian transformation (\ref{H_tun_chi}) is equivalent to a simple rotation of the coupling constants $\lambda_{i\alpha} \to \lambda_{i\alpha} e^\frac{i \chi(t)}{2N}$ in the original theory (\ref{H_full}). Thus, the Kadanoff-Baym equations (\ref{eqKBt},\ref{eqKBtp}) describe the valid saddle-point for the partition function (\ref{Z_chi}) up to the redefinition of the coupling constants $\lambda_i$. Indeed, the current can be deduced from the tunneling part of the effective action
\begin{align} \nonumber 
S_{tun}(\chi) =& i \sqrt{NM} \lambda^2 \! \sum_{ss'=\pm} \int_0^{+\infty} \!\! dt dt'  ss'  e^{\frac{i (s \chi(t)-s'\chi(t'))}{2N}}  \\ & \times G_{ss'}(t,t') Q_{s'\!s}(t'\!,t). \label{Stun}
\end{align}
Here the Green's functions $G_{ss'}$ and $Q_{ss'}$ describe the saddle-point of the SYK-bath system and are found from the equations (\ref{eqKBt}-\ref{Qgl}), where $s =\pm$ denotes the forward and backward branch of the Keldysh contour. Accordingly, the counting field $\chi$ is defined on the Keldysh contour as $\chi_s(t)=s\chi(t)$. Leaving the detailed derivation of the full effective action of the SYK-bath coupled system for the Appendix \ref{app:KB}, we proceed to the tunneling current 

Applying the prescription (\ref{I}), we derive the expectation value of current as a function of the measurement time $t_{\rm m}$:
\begin{align} \nonumber
    \mathcal{I}=& -\frac{\sqrt{p}\lambda^2}{2 t_{\rm m}} \sum_{ss'} \int_0^{t_{\rm m}} \!\! dt \int_0^{+\infty} \!\! dt' \Big(  G_{ss'}(t,t') s' Q_{s'\!s}(t'\!,t) \\ \nonumber &- Q_{s'\!s}(t,t') s \, G_{ss'}(t'\!,t)  \Big) \\ \nonumber =& -\frac{\sqrt{p}\lambda^2}{2 t_{\rm m}} \int_0^{t_{\rm m}} \!\! dt \int_0^{+\infty} \!\! dt' \, \mathrm{tr} \Big(  \sigma^x \hat{G}(t,t') \hat{Q}(t'\!,t) \\&- \sigma^x \hat{Q}(t,t') \hat{G}(t'\!,t)  \Big),   \label{It}
\end{align}
where
\begin{align}
\hat{G}=\begin{pmatrix}
G_R & G_K \\ 0 & G_A
\end{pmatrix}, \quad \hat{Q}=\begin{pmatrix}
Q_R & Q_K \\ 0 & Q_A
\end{pmatrix}
\end{align}
are the Green's functions of the SYK quantum dot and the cool-bath set by the equations (\ref{eqKBt}-\ref{Qgl}) and transformed to the $R,A,K$ basis according to the rules (\ref{GRgl}-\ref{GKgl}) \footnote{
In equilibrium the fluctuation dissipation relation holds $G_K(\omega)=-2\pi i (1-2 n_{SY\!K}(\omega)) \nu_{SY\!K}(\omega)$, $Q_K(\omega)=-2\pi i (1-2 n_{res}(\omega)) \nu_{res}(\omega)$, where $n_{SY\!K}$ and $n_{res}$ are the Fermi distribution functions and  $\nu_{SY\!K}=-\frac{1}{\pi} \mathrm{Im} G_R$ and $\nu_{res}=-\frac{1}{\pi} \mathrm{Im} Q_R$ are the densities of states. Substituting those to Eq. (\ref{It}), one gets a familiar Fermi golden rule formula for the tunneling current  \cite{Kamenev2005Course}: \\
$
\mathcal{I} = 2\pi \sqrt{p}\lambda^2 \! \int \! d\omega \, \nu_{SY\!K}(\omega) \nu_{res}(\omega) (n_{SY\!K}(\omega) - n_{res}(\omega))
$.}. 
From here, the dynamics of the tunneling current is given by
\begin{align}
\langle\dot{\mathcal{Q}}(t)\rangle = & -\frac{\sqrt{p}\lambda^2}{2} \theta(t) \int_0^t \!\! dt' \mathcal{J}(t,t'), \label{dQdt_res} \\ \nonumber
\mathcal{J}(t,t') = & \, G_R(t,t')Q_K(t'\!,t) - Q_K(t,t')G_A(t'\!,t) \\ &- Q_R(t,t')G_K(t'\!,t) +  G_K(t,t')Q_A(t'\!,t).
\end{align}

Time dependence of the tunneling current is shown in Fig. \ref{fig:dQdt}. The current grows initially as a power law, reaches the maximum value, and decays exponentially to zero consistently with the discharging process of the SYK quantum dot. With intention to mark the lifetime of the effect we extract the half-life -- the time in which the current is decreased in half of its maximum value. Varying the initial temperature $T$ of the SYK quantum dot, we show the current's half-life for several coupling strengths in Fig. \ref{fig:t05} (\textit{left}).  
The stronger the coupling, the shorter the half-life of the tunneling current. Oppositely, the half-life increases with the initial temperature rise. For the temperatures $T \lesssim 0.4 J$ the tunneling current half-life growths linearly in $T$.

To check if the $T$-linear current's half-life is specific for the SYK state, we substitute the SYK model with the one-body random Hamiltonian (\ref{H_bath}), often refereed to as the SYK$_2$ model, the same that describes the reservoir. This model has a typical Fermi liquid Green's function $G_R(t) \propto 1/t$ in the long time limit $J t \gg 1$,  which makes it legitimate to build the SYK nFl/Fermi liquid comparison. Matching the tunneling current half-life for the SYK vs SYK$_2$ model in Fig. \ref{fig:t05} (\textit{right}), we ascertain that their temperature dependencies are drastically different. The current's half-life in the system of the SYK$_2$ quantum dot coupled to the cold bath increases as $T^2$ at low temperatures, which discerns it from the SYK model cooling protocol displaying the linear in temperature increase.

Duration of the tunneling event in our system is defined by the tunneling contact resistance, similarly to an exponentially relaxing capacitor discharge. As such, our results resemble the prominent resistivity predictions for strange metals  $\rho_{SM} \sim T$ \cite{Varma2002Singular, Son2017Strongly, Chowdhury2018Translationally}  and Fermi liquid $\rho_{FL}\sim T^2$. 

\section{Conclusion}\label{sec:conclusion}

The Sachdev-Ye-Kitaev model quench-coupled with a cold bath has been a subject of close attention aiming to simulate evaporation of a black hole \cite{Almheiri2019Universal, Zhang2019Evaporation}. At the same time, both connecting the system to the environment and its further characterization are inherent for realization proposals of the SYK model in condensed matter systems \cite{Pikulin2017Black, Chen2018Quantum, Chew2017Approximating, Danshita2016Creating, Wei2021Optical}. In this manuscript, we consider a quantum dot described by the complex SYK model at finite temperature instantaneously coupled to a zero temperature reservoir. Analyzing the dynamical spectral function of the SYK quantum dot at charge neutrality, we show that the considered quench protocol preserves the SYK non-Fermi liquid state for the energies $\delta\omega\ll \omega\ll J$. Here the lower bound $\delta\omega$ is suppressed in the weak tunneling limit. Further, we put an initial electrochemical potential in the quantum dot and compute the tunneling current dynamics due to discharging of the dot. The tunneling current half-life shows distinct temperature dependencies for different systems that are being cooled down. In case of the SYK quantum dot, the half-life increases linearly in the initial temperature $T$, while for the Fermi liquid the increase is $\propto T^2$. Therefore, this temperature dependence of the tunneling current half-life provides a distinguishing feature for the disordered quantum dot exhibiting the SYK nFl phase against more common Fermi liquid behavior.



\acknowledgments 

We are grateful to Vladislav Kurilovich for the valuable comments on our results. We also have benefited from discussions with E. Yaraie and M. Kiselev. This research was supported in part by the Netherlands Organization for Scientific Research/Ministry of Science and Education (NWO/OCW), by the European Research Council (ERC) under the European Union’s Horizon 2020 research and innovation programme. 

\begin{widetext}

\appendix

\section{Derivation of the Kadanoff-Baym equations from the SYK saddle-point}\label{app:KB}

Here we derive the Kadanoff-Baym equations for the SYK quantum dot coupled to a cool-bath by a quench.

\subsection{Saddle-point equations}\label{app:SP}

We perform the disorder average of with the Hamiltonian (\ref{H_full}), pursuing \cite{Son2017Strongly, Can2019Charge}. The effective action  can be written in terms of bilocal fields $G_{s'\!s}(t'\!,t)=i N^{-1} \sum_i \bar{c}_{is}(t) c_{is}(t')$, $Q_{s'\!s}(t'\!,t)=i M^{-1} \sum_\alpha \bar{\psi}_{\alpha s}(t) \psi_{\alpha s}(t')$ and $\Sigma_{ss'}(t,t')$, $\Pi_{ss'}(t,t')$ as the corresponding Lagrange multipliers
\begin{align} \nonumber
S=&- i N \mathrm{tr}\! \ln\!\Big[\sigma^z_{ss'} \delta(t-t')\, \left(i \partial_t+\mu\right) - \Sigma_{ss'}(t,t')\Big]- i N \sum_{ss'} \! \int \!\! dt dt' \! \left(\!\Sigma_{ss'}(t,t')G_{s'\!s}(t'\!,t)-\frac{ss'J^2}{4} G_{ss'}(t,t')^2 G_{s'\!s}(t'\!,t)^2\!\right)  \\ \nonumber 
&- i M \mathrm{tr}\! \ln\!\Big[\sigma^z_{ss'} \delta(t-t')\, i \partial_t - \Pi_{ss'}(t,t')\Big]- i M \sum_{ss'} \! \int \!\! dt dt' \! \left(\!\Pi_{ss'}(t,t')Q_{s'\!s}(t'\!,t)-\frac{ss'\xi^2}{2} Q_{ss'}(t,t') Q_{s'\!s}(t'\!,t)\!\right) \\
&+ i\sqrt{NM} \sum_{ss'} \! \int \!\! dt dt'  ss' \lambda^2 \theta(t) \theta(t') G_{ss'}(t,t') Q_{s'\!s}(t'\!,t) .  \label{S_eff}
\end{align}
where $s=\pm$ denotes forward and backward branches of the Keldysh time contour \cite{Kamenev2005Course}. 
In the large $N$, $M$ limit, the saddle-point equations are
\begin{align}
&\Sigma_{ss'}(t,t')=J^2 G_{ss'}(t,t')^2 G_{s'\!s}(t'\!,t)+\sqrt{p} \, \lambda^2 \theta(t) \theta(t') Q_{ss'}(t,t'), \label{eqSigma}\\
&\Pi_{ss'}(t,t')=\xi^2 Q_{ss'}(t,t')+\frac{\lambda^2}{\sqrt{p}}  \, \theta(t) \theta(t') G_{ss'}(t,t'), \label{eqP}\\
&\sum_r \! \int_{-\infty}^{+\infty} \!\!\! du \Big( \sigma^z_{sr} \delta(t-u) \left(i\partial_t+\mu\right) - sr \, \Sigma_{sr}(t,u)\Big)G_{rs'}(u,t')=\delta_{ss'}\delta(t-t'),\label{eqG}\\
&\sum_r \! \int_{-\infty}^{+\infty} \!\!\! du \Big( \sigma^z_{sr} \delta(t-u) i\partial_t - sr \, \Pi_{sr}(t,u)\Big)Q_{rs'}(u,t')=\delta_{ss'}\delta(t-t'),\label{eqQ}
\end{align} 
where $p=M/N$ is the mode ratio.

Following Ref. \cite{Eberlein2017Quantum}, we derive the self-consistent Kadanoff-Baym equations considering $s,s'=\pm,\mp$ components of Eqs. (\ref{eqG}, \ref{eqQ}):
\begin{align}
\left(i \partial_t+\mu\right) G^\gtrless (t,t') & = \int_{-\infty}^{+\infty}\!\!du \Big( \Sigma_R(t,u) G^\gtrless (u,t') + \Sigma^\gtrless (t,u) G_A(u,t')\Big), \label{sc_KB1}\\
\left(-i \partial_{t'}+\mu\right) G^\gtrless (t,t') & = \int_{-\infty}^{+\infty}\!\!du \Big( G_R(t,u) \Sigma^\gtrless (u,t') + G^\gtrless (t,u) \Sigma_A(u,t')\Big), \label{sc_KB2} \\
i \partial_t Q^\gtrless (t,t') & = \int_{-\infty}^{+\infty}\!\!du \Big( \Pi_R(t,u) Q^\gtrless (u,t') + \Pi^\gtrless (t,u) Q_A(u,t')\Big), \label{sc_KB3}\\
-i \partial_{t'} Q^\gtrless (t,t') & = \int_{-\infty}^{+\infty}\!\!du \Big( Q_R(t,u) \Pi^\gtrless (u,t') + Q^\gtrless (t,u) \Pi_A(u,t')\Big), \label{sc_KB4}
\end{align}
where the self-energies are
\begin{align}
\Sigma^\gtrless (t,t') & = J^2 G^\gtrless(t,t')^2 G^\lessgtr(t'\!,t) + \sqrt{p} \, \lambda^2 \theta(t) \theta(t') Q^\gtrless(t,t'), \label{sc_SigmaKB} \\
\Pi^\gtrless (t,t') & = \xi^2 Q^\gtrless(t,t') + \frac{\lambda^2}{\sqrt{p}} \, \theta(t) \theta(t') G^\gtrless(t,t'). \label{sc_PiKB}
\end{align}

\subsection{Reservoir as an external potential}\label{app:reservoir}

Since we assume the reservoir to be large enough $p \gg 1$, it can be considered as a closed dynamic background to the SYK subsystem
\begin{align}
&\int_{-\infty}^{+\infty} \!\!\! du \Big( \delta(t-u) i\partial_t - \xi^2 \hat{Q}(t,u)\Big)\hat{Q}(u,t')=\delta(t-t')\label{eqQmat}
\end{align}
describing a decoupled random free fermion in equilibrium.
Here we perform a rotation towards retarded, advanced, and Keldysh basis
$$\hat{Q}=\begin{pmatrix}
Q_R & Q_K \\ 0 & Q_A
\end{pmatrix} =L\sigma^z\begin{pmatrix}
Q_{++} & Q_{+-} \\ Q_{-+} & Q_{--}
\end{pmatrix}L^\dag, \quad L=\frac{1}{\sqrt{2}}\begin{pmatrix}
1 & -1 \\ 1 & 1
\end{pmatrix}.$$
The retarded Green's function is found from
$$ \left(\omega-\xi^2 Q_R(\omega)\right)Q_R(\omega)=1 \quad \Rightarrow \quad Q_R(\omega)=\frac{\omega}{2\xi^2}-\frac{i}{\xi}\sqrt{1-\frac{\omega^2}{4\xi^2}} = \frac{2}{\omega+2 i \xi \sqrt{1-\left(\omega/2 \xi\right)^2}}, $$
where the spectral function obeys the semicircle law $\rho(\omega)=-2 \, \mathrm{Im}Q_R(\omega)=\dfrac{2}{\xi}\, \mathrm{Re}\sqrt{1-\dfrac{\omega^2}{4 \xi^2}}$. 
Let's derive the time representation of $Q_R$:
\begin{align}
Q_R(t,t')=Q_A(t'\!,t)^*=\int_{-\infty}^{+\infty}\,\, \frac{d\omega}{2\pi} \,e^{-i\omega(t-t')} Q_R(\omega)=-\lim_{\delta\to 0^+}\int_{-\infty}^{+\infty}\,\, \frac{d\omega}{2\pi} \,e^{-i\omega(t-t')}e^{\delta (t-t')} \frac{1}{2\xi^2}\sqrt{\left(\omega+i\delta\right)^2-4\xi^2}. \label{QR_FT}
\end{align}
Here the branch cut is in the lower half plane, so we close the contour correspondingly for $t-t'>0$. Since there are no poles in the lower half plane, we shrink the contour to the anticlockwise traverse around the branch cut. Note that an additional phase is acquired when crossing the branch cut $\sqrt{\omega^2-4\xi^2}\to e^{\frac{1}{2}\ln \left(\omega^2+4\xi^2 \right)+i\pi}=e^{i\pi}\sqrt{\omega^2-4\xi^2}$. 
Therefore, we get 
\begin{align}
Q_R(t,t')=-\theta(t-t')\frac{1-e^{i\pi}}{4\pi\xi^2}\int_{-2\xi}^{2\xi}\,\, d\omega \,e^{-i\omega(t-t')} \sqrt{\omega^2-4\xi^2}=-i \theta(t-t')\, \frac{J_1\!\left(2\xi (t-t')\right)}{\xi(t-t')}, \label{QR}
\end{align} 
where $J_1$ is the first Bessel function of the first kind.
The Keldysh component at zero temperature is
\begin{align} \nonumber 
Q_K(t,t')=&\int_{-\infty}^{+\infty}\,\, \frac{d\omega}{2\pi} \,e^{-i\omega(t-t')} Q_K(\omega) = \int_{-\infty}^{+\infty}\,\, \frac{d\omega}{2\pi} \,e^{-i\omega(t-t')} 2 i\, \mathrm{sgn}(\omega) \mathrm{Im} Q_R(\omega) \\ =&-\frac{i}{2\pi \xi^2}\int_{-2\xi}^{2\xi}\,\, d\omega \,e^{-i\omega(t-t')} \, \mathrm{sgn}(\omega)   \sqrt{4\xi^2-\omega^2} = -\frac{\mathbf{H}_1\!\left(2\xi (t-t')\right)}{\xi(t-t')}, \label{QK}
\end{align} 
where $\mathbf{H}_1$ is the first Struve function. 

\subsection{Dynamics of the SYK subsystem}\label{app:SYK}
In the large $p$ limit, the dynamics of the SYK subsystem is described by Eqs. (\ref{sc_KB1},\ref{sc_KB2},\ref{sc_SigmaKB}), where the reservoir Green's function $Q(t-t')$ enters the SYK self-energy (\ref{sc_SigmaKB}) as the external potential derived in Section \ref{app:reservoir}. Thereby, the Kadanoff-Baym equations simplify to
\begin{align}
\left(i \partial_t+\mu\right) G^\gtrless (t,t') & = \int_{-\infty}^{+\infty}\!\!du \Big( \Sigma_R(t,u) G^\gtrless (u,t') + \Sigma^\gtrless (t,u) G_A(u,t')\Big), \label{KB1}\\
\left(-i \partial_{t'}+\mu\right) G^\gtrless (t,t') & = \int_{-\infty}^{+\infty}\!\!du \Big( G_R(t,u) \Sigma^\gtrless (u,t') + G^\gtrless (t,u) \Sigma_A(u,t')\Big), \label{KB2}
\end{align}
with the self-energy (\ref{eqSigma})
\begin{align}
\Sigma^\gtrless (t,t') & = J^2 G^\gtrless(t,t')^2 G^\lessgtr(t'\!,t)+\sqrt{p} \, \lambda^2 \theta(t) \theta(t') Q^\gtrless(t,t'), \label{SigmaKB}\\
Q^\gtrless(t,t')&=-\frac{1}{2\xi(t-t')}\Big(\mathbf{H}_1\!\left(2\xi (t-t')\right) \pm i J_1\!\left(2\xi (t-t')\right) \Big).
\end{align}
Here we introduced \cite{Kamenev2005Course} $G^>(t,t')\equiv G_{-+}(t,t')$, $G^<(t,t') \equiv G_{+-}(t,t')$, $\Sigma^>(t,t')\equiv \Sigma_{-+}(t,t')$, $\Sigma^<(t,t') \equiv \Sigma_{+-}(t,t')$ and account for \begin{align}
G_{++}(t,t') & = \theta(t-t') G^>(t,t') + \theta(t'\!-t) G^<(t,t'), \label{Gpp} \\
G_{--}(t,t') & = \theta(t'\!-t) G^>(t,t') + \theta(t-t') G^<(t,t'), \label{Gmm}\\
\Sigma_{++}(t,t') & = \theta(t-t') \Sigma^>(t,t') + \theta(t'\!-t) \Sigma^<(t,t'), \label{Spp} \\
\Sigma_{--}(t,t') & = \theta(t'\!-t) \Sigma^>(t,t') + \theta(t-t') \Sigma^<(t,t'). \label{Smm}
\end{align}   
The retarded, advanced, and Keldysh components are expressed in terms of $>$  and $<$ as
\begin{align}
G_R(t,t') & = \theta(t-t') \Big(G^>(t,t') - G^<(t,t')\Big), \label{GR}\\
G_A(t,t') & = -\theta(t'\!-t)  \Big(G^>(t,t') - G^<(t,t')\Big), \label{GA}\\
G_K(t,t') & = G^>(t,t') + G^<(t,t'), \label{GK} \\
\Sigma_R(t,t') & = \theta(t-t') \Big(\Sigma^>(t,t') - \Sigma^<(t,t')\Big), \label{SR}\\
\Sigma_A(t,t') & = -\theta(t'\!-t)  \Big(\Sigma^>(t,t') - \Sigma^<(t,t')\Big), \label{SA}\\
\Sigma_K(t,t') & = \Sigma^>(t,t') + \Sigma^<(t,t') \label{SK}. 
\end{align}

\end{widetext}
 
\bibliography{refs} 

\begin{thebibliography}{45}%
\makeatletter
\providecommand \@ifxundefined [1]{%
 \@ifx{#1\undefined}
}%
\providecommand \@ifnum [1]{%
 \ifnum #1\expandafter \@firstoftwo
 \else \expandafter \@secondoftwo
 \fi
}%
\providecommand \@ifx [1]{%
 \ifx #1\expandafter \@firstoftwo
 \else \expandafter \@secondoftwo
 \fi
}%
\providecommand \natexlab [1]{#1}%
\providecommand \enquote  [1]{``#1''}%
\providecommand \bibnamefont  [1]{#1}%
\providecommand \bibfnamefont [1]{#1}%
\providecommand \citenamefont [1]{#1}%
\providecommand \href@noop [0]{\@secondoftwo}%
\providecommand \href [0]{\begingroup \@sanitize@url \@href}%
\providecommand \@href[1]{\@@startlink{#1}\@@href}%
\providecommand \@@href[1]{\endgroup#1\@@endlink}%
\providecommand \@sanitize@url [0]{\catcode `\\12\catcode `\$12\catcode
  `\&12\catcode `\#12\catcode `\^12\catcode `\_12\catcode `\%12\relax}%
\providecommand \@@startlink[1]{}%
\providecommand \@@endlink[0]{}%
\providecommand \url  [0]{\begingroup\@sanitize@url \@url }%
\providecommand \@url [1]{\endgroup\@href {#1}{\urlprefix }}%
\providecommand \urlprefix  [0]{URL }%
\providecommand \Eprint [0]{\href }%
\providecommand \doibase [0]{http://dx.doi.org/}%
\providecommand \selectlanguage [0]{\@gobble}%
\providecommand \bibinfo  [0]{\@secondoftwo}%
\providecommand \bibfield  [0]{\@secondoftwo}%
\providecommand \translation [1]{[#1]}%
\providecommand \BibitemOpen [0]{}%
\providecommand \bibitemStop [0]{}%
\providecommand \bibitemNoStop [0]{.\EOS\space}%
\providecommand \EOS [0]{\spacefactor3000\relax}%
\providecommand \BibitemShut  [1]{\csname bibitem#1\endcsname}%
\let\auto@bib@innerbib\@empty
\bibitem [{\citenamefont {Kitaev}(2015)}]{Kitaev2015Simple}%
  \BibitemOpen
  \bibfield  {author} {\bibinfo {author} {\bibfnamefont {A.}~\bibnamefont
  {Kitaev}},\ }\href@noop {} {\emph {\bibinfo {title} {A simple model of
  quantum holography}}}\ (\bibinfo  {publisher} {KITP Program: Entanglement in
  Strongly-Correlated Quantum Matter},\ \bibinfo {year} {2015})\BibitemShut
  {NoStop}%
\bibitem [{\citenamefont {Sachdev}\ and\ \citenamefont
  {Ye}(1993)}]{Sachdev1993Gapless}%
  \BibitemOpen
  \bibfield  {author} {\bibinfo {author} {\bibfnamefont {S.}~\bibnamefont
  {Sachdev}}\ and\ \bibinfo {author} {\bibfnamefont {J.}~\bibnamefont {Ye}},\
  }\href {\doibase 10.1103/PhysRevLett.70.3339} {\bibfield  {journal} {\bibinfo
   {journal} {Phys. Rev. Lett.}\ }\textbf {\bibinfo {volume} {70}},\ \bibinfo
  {pages} {3339} (\bibinfo {year} {1993})}\BibitemShut {NoStop}%
\bibitem [{\citenamefont {Almheiri}\ \emph
  {et~al.}(2019{\natexlab{a}})\citenamefont {Almheiri}, \citenamefont
  {Milekhin},\ and\ \citenamefont {Swingle}}]{Almheiri2019Universal}%
  \BibitemOpen
  \bibfield  {author} {\bibinfo {author} {\bibfnamefont {A.}~\bibnamefont
  {Almheiri}}, \bibinfo {author} {\bibfnamefont {A.}~\bibnamefont {Milekhin}},
  \ and\ \bibinfo {author} {\bibfnamefont {B.}~\bibnamefont {Swingle}},\
  }\href@noop {} {\  (\bibinfo {year} {2019}{\natexlab{a}})},\ \Eprint
  {http://arxiv.org/abs/1912.04912} {arXiv:1912.04912 [hep-th]} \BibitemShut
  {NoStop}%
\bibitem [{\citenamefont {Zhang}(2019)}]{Zhang2019Evaporation}%
  \BibitemOpen
  \bibfield  {author} {\bibinfo {author} {\bibfnamefont {P.}~\bibnamefont
  {Zhang}},\ }\href {\doibase 10.1103/PhysRevB.100.245104} {\bibfield
  {journal} {\bibinfo  {journal} {Phys. Rev. B}\ }\textbf {\bibinfo {volume}
  {100}},\ \bibinfo {pages} {245104} (\bibinfo {year} {2019})}\BibitemShut
  {NoStop}%
\bibitem [{\citenamefont {Hawking}(1974)}]{Hawking1974Black}%
  \BibitemOpen
  \bibfield  {author} {\bibinfo {author} {\bibfnamefont {S.~W.}\ \bibnamefont
  {Hawking}},\ }\href {\doibase 10.1038/248030a0} {\bibfield  {journal}
  {\bibinfo  {journal} {Nature}\ }\textbf {\bibinfo {volume} {248}},\ \bibinfo
  {pages} {30} (\bibinfo {year} {1974})}\BibitemShut {NoStop}%
\bibitem [{\citenamefont {Hawking}(1976)}]{Hawking1976Breakdown}%
  \BibitemOpen
  \bibfield  {author} {\bibinfo {author} {\bibfnamefont {S.~W.}\ \bibnamefont
  {Hawking}},\ }\href {\doibase 10.1103/PhysRevD.14.2460} {\bibfield  {journal}
  {\bibinfo  {journal} {Phys. Rev. D}\ }\textbf {\bibinfo {volume} {14}},\
  \bibinfo {pages} {2460} (\bibinfo {year} {1976})}\BibitemShut {NoStop}%
\bibitem [{\citenamefont {Page}(1993)}]{Page1993Information}%
  \BibitemOpen
  \bibfield  {author} {\bibinfo {author} {\bibfnamefont {D.~N.}\ \bibnamefont
  {Page}},\ }\href {\doibase 10.1103/PhysRevLett.71.3743} {\bibfield  {journal}
  {\bibinfo  {journal} {Phys. Rev. Lett.}\ }\textbf {\bibinfo {volume} {71}},\
  \bibinfo {pages} {3743} (\bibinfo {year} {1993})}\BibitemShut {NoStop}%
\bibitem [{\citenamefont {Almheiri}\ \emph
  {et~al.}(2019{\natexlab{b}})\citenamefont {Almheiri}, \citenamefont
  {Engelhardt}, \citenamefont {Marolf},\ and\ \citenamefont
  {Maxfield}}]{Almheiri2019Entropy}%
  \BibitemOpen
  \bibfield  {author} {\bibinfo {author} {\bibfnamefont {A.}~\bibnamefont
  {Almheiri}}, \bibinfo {author} {\bibfnamefont {N.}~\bibnamefont
  {Engelhardt}}, \bibinfo {author} {\bibfnamefont {D.}~\bibnamefont {Marolf}},
  \ and\ \bibinfo {author} {\bibfnamefont {H.}~\bibnamefont {Maxfield}},\
  }\href {\doibase 10.1007/JHEP12(2019)063} {\bibfield  {journal} {\bibinfo
  {journal} {J. High Energy Phys.}\ }\textbf {\bibinfo {volume} {2019}},\
  \bibinfo {pages} {63} (\bibinfo {year} {2019}{\natexlab{b}})}\BibitemShut
  {NoStop}%
\bibitem [{\citenamefont {Almheiri}\ \emph
  {et~al.}(2019{\natexlab{c}})\citenamefont {Almheiri}, \citenamefont
  {Mahajan},\ and\ \citenamefont {Maldacena}}]{Almheiri2019Islands}%
  \BibitemOpen
  \bibfield  {author} {\bibinfo {author} {\bibfnamefont {A.}~\bibnamefont
  {Almheiri}}, \bibinfo {author} {\bibfnamefont {R.}~\bibnamefont {Mahajan}}, \
  and\ \bibinfo {author} {\bibfnamefont {J.}~\bibnamefont {Maldacena}},\
  }\href@noop {} {\  (\bibinfo {year} {2019}{\natexlab{c}})},\ \Eprint
  {http://arxiv.org/abs/1910.11077} {arXiv:1910.11077 [hep-th]} \BibitemShut
  {NoStop}%
\bibitem [{\citenamefont {Pikulin}\ and\ \citenamefont
  {Franz}(2017)}]{Pikulin2017Black}%
  \BibitemOpen
  \bibfield  {author} {\bibinfo {author} {\bibfnamefont {D.~I.}\ \bibnamefont
  {Pikulin}}\ and\ \bibinfo {author} {\bibfnamefont {M.}~\bibnamefont
  {Franz}},\ }\href {\doibase 10.1103/PhysRevX.7.031006} {\bibfield  {journal}
  {\bibinfo  {journal} {Phys. Rev. X}\ }\textbf {\bibinfo {volume} {7}},\
  \bibinfo {pages} {031006} (\bibinfo {year} {2017})}\BibitemShut {NoStop}%
\bibitem [{\citenamefont {Chew}\ \emph {et~al.}(2017)\citenamefont {Chew},
  \citenamefont {Essin},\ and\ \citenamefont {Alicea}}]{Chew2017Approximating}%
  \BibitemOpen
  \bibfield  {author} {\bibinfo {author} {\bibfnamefont {A.}~\bibnamefont
  {Chew}}, \bibinfo {author} {\bibfnamefont {A.}~\bibnamefont {Essin}}, \ and\
  \bibinfo {author} {\bibfnamefont {J.}~\bibnamefont {Alicea}},\ }\href
  {\doibase 10.1103/PhysRevB.96.121119} {\bibfield  {journal} {\bibinfo
  {journal} {Phys. Rev. B}\ }\textbf {\bibinfo {volume} {96}},\ \bibinfo
  {pages} {121119(R)} (\bibinfo {year} {2017})}\BibitemShut {NoStop}%
\bibitem [{\citenamefont {Danshita}\ \emph {et~al.}(2017)\citenamefont
  {Danshita}, \citenamefont {Hanada},\ and\ \citenamefont
  {Tezuka}}]{Danshita2016Creating}%
  \BibitemOpen
  \bibfield  {author} {\bibinfo {author} {\bibfnamefont {I.}~\bibnamefont
  {Danshita}}, \bibinfo {author} {\bibfnamefont {M.}~\bibnamefont {Hanada}}, \
  and\ \bibinfo {author} {\bibfnamefont {M.}~\bibnamefont {Tezuka}},\ }\href
  {\doibase 10.1093/ptep/ptx108} {\bibfield  {journal} {\bibinfo  {journal}
  {PTEP}\ }\textbf {\bibinfo {volume} {2017}},\ \bibinfo {pages} {083I01}
  (\bibinfo {year} {2017})}\BibitemShut {NoStop}%
\bibitem [{\citenamefont {Wei}\ and\ \citenamefont
  {Sedrakyan}(2021)}]{Wei2021Optical}%
  \BibitemOpen
  \bibfield  {author} {\bibinfo {author} {\bibfnamefont {C.}~\bibnamefont
  {Wei}}\ and\ \bibinfo {author} {\bibfnamefont {T.~A.}\ \bibnamefont
  {Sedrakyan}},\ }\href {\doibase 10.1103/PhysRevA.103.013323} {\bibfield
  {journal} {\bibinfo  {journal} {Phys. Rev. A}\ }\textbf {\bibinfo {volume}
  {103}},\ \bibinfo {pages} {013323} (\bibinfo {year} {2021})}\BibitemShut
  {NoStop}%
\bibitem [{\citenamefont {Chen}\ \emph {et~al.}(2018)\citenamefont {Chen},
  \citenamefont {Ilan}, \citenamefont {de~Juan}, \citenamefont {Pikulin},\ and\
  \citenamefont {Franz}}]{Chen2018Quantum}%
  \BibitemOpen
  \bibfield  {author} {\bibinfo {author} {\bibfnamefont {A.}~\bibnamefont
  {Chen}}, \bibinfo {author} {\bibfnamefont {R.}~\bibnamefont {Ilan}}, \bibinfo
  {author} {\bibfnamefont {F.}~\bibnamefont {de~Juan}}, \bibinfo {author}
  {\bibfnamefont {D.~I.}\ \bibnamefont {Pikulin}}, \ and\ \bibinfo {author}
  {\bibfnamefont {M.}~\bibnamefont {Franz}},\ }\href {\doibase
  10.1103/PhysRevLett.121.036403} {\bibfield  {journal} {\bibinfo  {journal}
  {Phys. Rev. Lett.}\ }\textbf {\bibinfo {volume} {121}},\ \bibinfo {pages}
  {036403} (\bibinfo {year} {2018})}\BibitemShut {NoStop}%
\bibitem [{\citenamefont {Garc\'{\i}a-\'Alvarez}\ \emph
  {et~al.}(2017)\citenamefont {Garc\'{\i}a-\'Alvarez}, \citenamefont
  {Egusquiza}, \citenamefont {Lamata}, \citenamefont {del Campo}, \citenamefont
  {Sonner},\ and\ \citenamefont {Solano}}]{Garcia2017Digital}%
  \BibitemOpen
  \bibfield  {author} {\bibinfo {author} {\bibfnamefont {L.}~\bibnamefont
  {Garc\'{\i}a-\'Alvarez}}, \bibinfo {author} {\bibfnamefont {I.~L.}\
  \bibnamefont {Egusquiza}}, \bibinfo {author} {\bibfnamefont {L.}~\bibnamefont
  {Lamata}}, \bibinfo {author} {\bibfnamefont {A.}~\bibnamefont {del Campo}},
  \bibinfo {author} {\bibfnamefont {J.}~\bibnamefont {Sonner}}, \ and\ \bibinfo
  {author} {\bibfnamefont {E.}~\bibnamefont {Solano}},\ }\href {\doibase
  10.1103/PhysRevLett.119.040501} {\bibfield  {journal} {\bibinfo  {journal}
  {Phys. Rev. Lett.}\ }\textbf {\bibinfo {volume} {119}},\ \bibinfo {pages}
  {040501} (\bibinfo {year} {2017})}\BibitemShut {NoStop}%
\bibitem [{\citenamefont {Luo}\ \emph {et~al.}(2019)\citenamefont {Luo},
  \citenamefont {You}, \citenamefont {Li}, \citenamefont {Jian}, \citenamefont
  {Lu}, \citenamefont {Xu}, \citenamefont {Zeng},\ and\ \citenamefont
  {Laflamme}}]{Luo2019Quantum}%
  \BibitemOpen
  \bibfield  {author} {\bibinfo {author} {\bibfnamefont {Z.}~\bibnamefont
  {Luo}}, \bibinfo {author} {\bibfnamefont {Y.-Z.}\ \bibnamefont {You}},
  \bibinfo {author} {\bibfnamefont {J.}~\bibnamefont {Li}}, \bibinfo {author}
  {\bibfnamefont {C.-M.}\ \bibnamefont {Jian}}, \bibinfo {author}
  {\bibfnamefont {D.}~\bibnamefont {Lu}}, \bibinfo {author} {\bibfnamefont
  {C.}~\bibnamefont {Xu}}, \bibinfo {author} {\bibfnamefont {B.}~\bibnamefont
  {Zeng}}, \ and\ \bibinfo {author} {\bibfnamefont {R.}~\bibnamefont
  {Laflamme}},\ }\href {\doibase 10.1038/s41534-019-0166-7} {\bibfield
  {journal} {\bibinfo  {journal} {npj Quantum Information}\ }\textbf {\bibinfo
  {volume} {5}},\ \bibinfo {pages} {53} (\bibinfo {year} {2019})}\BibitemShut
  {NoStop}%
\bibitem [{\citenamefont {Babbush}\ \emph {et~al.}(2019)\citenamefont
  {Babbush}, \citenamefont {Berry},\ and\ \citenamefont
  {Neven}}]{Babbush2019Quantum}%
  \BibitemOpen
  \bibfield  {author} {\bibinfo {author} {\bibfnamefont {R.}~\bibnamefont
  {Babbush}}, \bibinfo {author} {\bibfnamefont {D.~W.}\ \bibnamefont {Berry}},
  \ and\ \bibinfo {author} {\bibfnamefont {H.}~\bibnamefont {Neven}},\ }\href
  {\doibase 10.1103/PhysRevA.99.040301} {\bibfield  {journal} {\bibinfo
  {journal} {Phys. Rev. A}\ }\textbf {\bibinfo {volume} {99}},\ \bibinfo
  {pages} {040301(R)} (\bibinfo {year} {2019})}\BibitemShut {NoStop}%
\bibitem [{\citenamefont {Sachdev}(2015)}]{Sachdev2015Bekenstein}%
  \BibitemOpen
  \bibfield  {author} {\bibinfo {author} {\bibfnamefont {S.}~\bibnamefont
  {Sachdev}},\ }\href {\doibase 10.1103/PhysRevX.5.041025} {\bibfield
  {journal} {\bibinfo  {journal} {Phys. Rev. X}\ }\textbf {\bibinfo {volume}
  {5}},\ \bibinfo {pages} {041025} (\bibinfo {year} {2015})}\BibitemShut
  {NoStop}%
\bibitem [{\citenamefont {Gu}\ \emph {et~al.}(2020)\citenamefont {Gu},
  \citenamefont {Kitaev}, \citenamefont {Sachdev},\ and\ \citenamefont
  {Tarnopolsky}}]{Gu2019Notes}%
  \BibitemOpen
  \bibfield  {author} {\bibinfo {author} {\bibfnamefont {Y.}~\bibnamefont
  {Gu}}, \bibinfo {author} {\bibfnamefont {A.}~\bibnamefont {Kitaev}}, \bibinfo
  {author} {\bibfnamefont {S.}~\bibnamefont {Sachdev}}, \ and\ \bibinfo
  {author} {\bibfnamefont {G.}~\bibnamefont {Tarnopolsky}},\ }\href {\doibase
  10.1007/JHEP02(2020)157} {\bibfield  {journal} {\bibinfo  {journal} {J. High
  Energy Phys.}\ }\textbf {\bibinfo {volume} {2020}},\ \bibinfo {pages} {157}
  (\bibinfo {year} {2020})}\BibitemShut {NoStop}%
\bibitem [{\citenamefont {Gnezdilov}\ \emph {et~al.}(2018)\citenamefont
  {Gnezdilov}, \citenamefont {Hutasoit},\ and\ \citenamefont
  {Beenakker}}]{Gnezdilov2018Low}%
  \BibitemOpen
  \bibfield  {author} {\bibinfo {author} {\bibfnamefont {N.~V.}\ \bibnamefont
  {Gnezdilov}}, \bibinfo {author} {\bibfnamefont {J.~A.}\ \bibnamefont
  {Hutasoit}}, \ and\ \bibinfo {author} {\bibfnamefont {C.~W.~J.}\ \bibnamefont
  {Beenakker}},\ }\href {\doibase 10.1103/PhysRevB.98.081413} {\bibfield
  {journal} {\bibinfo  {journal} {Phys. Rev. B}\ }\textbf {\bibinfo {volume}
  {98}},\ \bibinfo {pages} {081413(R)} (\bibinfo {year} {2018})}\BibitemShut
  {NoStop}%
\bibitem [{\citenamefont {Can}\ \emph {et~al.}(2019)\citenamefont {Can},
  \citenamefont {Nica},\ and\ \citenamefont {Franz}}]{Can2019Charge}%
  \BibitemOpen
  \bibfield  {author} {\bibinfo {author} {\bibfnamefont {O.}~\bibnamefont
  {Can}}, \bibinfo {author} {\bibfnamefont {E.~M.}\ \bibnamefont {Nica}}, \
  and\ \bibinfo {author} {\bibfnamefont {M.}~\bibnamefont {Franz}},\ }\href
  {\doibase 10.1103/PhysRevB.99.045419} {\bibfield  {journal} {\bibinfo
  {journal} {Phys. Rev. B}\ }\textbf {\bibinfo {volume} {99}},\ \bibinfo
  {pages} {045419} (\bibinfo {year} {2019})}\BibitemShut {NoStop}%
\bibitem [{\citenamefont {Altland}\ \emph {et~al.}(2019)\citenamefont
  {Altland}, \citenamefont {Bagrets},\ and\ \citenamefont
  {Kamenev}}]{Altland2019Sachdev}%
  \BibitemOpen
  \bibfield  {author} {\bibinfo {author} {\bibfnamefont {A.}~\bibnamefont
  {Altland}}, \bibinfo {author} {\bibfnamefont {D.}~\bibnamefont {Bagrets}}, \
  and\ \bibinfo {author} {\bibfnamefont {A.}~\bibnamefont {Kamenev}},\ }\href
  {\doibase 10.1103/PhysRevLett.123.226801} {\bibfield  {journal} {\bibinfo
  {journal} {Phys. Rev. Lett.}\ }\textbf {\bibinfo {volume} {123}},\ \bibinfo
  {pages} {226801} (\bibinfo {year} {2019})}\BibitemShut {NoStop}%
\bibitem [{\citenamefont {Kruchkov}\ \emph {et~al.}(2020)\citenamefont
  {Kruchkov}, \citenamefont {Patel}, \citenamefont {Kim},\ and\ \citenamefont
  {Sachdev}}]{Kruchkov2020Thermoelectric}%
  \BibitemOpen
  \bibfield  {author} {\bibinfo {author} {\bibfnamefont {A.}~\bibnamefont
  {Kruchkov}}, \bibinfo {author} {\bibfnamefont {A.~A.}\ \bibnamefont {Patel}},
  \bibinfo {author} {\bibfnamefont {P.}~\bibnamefont {Kim}}, \ and\ \bibinfo
  {author} {\bibfnamefont {S.}~\bibnamefont {Sachdev}},\ }\href {\doibase
  10.1103/PhysRevB.101.205148} {\bibfield  {journal} {\bibinfo  {journal}
  {Phys. Rev. B}\ }\textbf {\bibinfo {volume} {101}},\ \bibinfo {pages}
  {205148} (\bibinfo {year} {2020})}\BibitemShut {NoStop}%
\bibitem [{\citenamefont {Pavlov}\ and\ \citenamefont
  {Kiselev}(2021)}]{Pavlov2020Quantum}%
  \BibitemOpen
  \bibfield  {author} {\bibinfo {author} {\bibfnamefont {A.~I.}\ \bibnamefont
  {Pavlov}}\ and\ \bibinfo {author} {\bibfnamefont {M.~N.}\ \bibnamefont
  {Kiselev}},\ }\href {\doibase 10.1103/PhysRevB.103.L201107} {\bibfield
  {journal} {\bibinfo  {journal} {Phys. Rev. B}\ }\textbf {\bibinfo {volume}
  {103}},\ \bibinfo {pages} {L201107} (\bibinfo {year} {2021})}\BibitemShut
  {NoStop}%
\bibitem [{\citenamefont
  {Khveshchenko}(2020{\natexlab{a}})}]{Khveshchenko2020SET}%
  \BibitemOpen
  \bibfield  {author} {\bibinfo {author} {\bibfnamefont {D.~V.}\ \bibnamefont
  {Khveshchenko}},\ }\href {\doibase 10.3952/physics.v60i3.4305} {\bibfield
  {journal} {\bibinfo  {journal} {Lithuanian Journal of Physics}\ }\textbf
  {\bibinfo {volume} {60}},\ \bibinfo {pages} {3} (\bibinfo {year}
  {2020}{\natexlab{a}})}\BibitemShut {NoStop}%
\bibitem [{\citenamefont
  {Khveshchenko}(2020{\natexlab{b}})}]{Khveshchenko2020Connecting}%
  \BibitemOpen
  \bibfield  {author} {\bibinfo {author} {\bibfnamefont {D.~V.}\ \bibnamefont
  {Khveshchenko}},\ }\href {\doibase 10.3390/condmat5020037} {\bibfield
  {journal} {\bibinfo  {journal} {Condensed Matter}\ }\textbf {\bibinfo
  {volume} {5}},\ \bibinfo {pages} {37} (\bibinfo {year}
  {2020}{\natexlab{b}})}\BibitemShut {NoStop}%
\bibitem [{\citenamefont {Kamenev}(2005)}]{Kamenev2005Course}%
  \BibitemOpen
  \bibfield  {author} {\bibinfo {author} {\bibfnamefont {A.}~\bibnamefont
  {Kamenev}},\ }in\ \href {\doibase
  https://doi.org/10.1016/S0924-8099(05)80045-9} {\emph {\bibinfo {booktitle}
  {Nanophysics: Coherence and Transport}}},\ \bibinfo {series} {Les Houches},
  Vol.~\bibinfo {volume} {81},\ \bibinfo {editor} {edited by\ \bibinfo {editor}
  {\bibfnamefont {H.}~\bibnamefont {Bouchiat}}, \bibinfo {editor}
  {\bibfnamefont {Y.}~\bibnamefont {Gefen}}, \bibinfo {editor} {\bibfnamefont
  {S.}~\bibnamefont {Guéron}}, \bibinfo {editor} {\bibfnamefont
  {G.}~\bibnamefont {Montambaux}}, \ and\ \bibinfo {editor} {\bibfnamefont
  {J.}~\bibnamefont {Dalibard}}}\ (\bibinfo  {publisher} {Elsevier},\ \bibinfo
  {year} {2005})\ pp.\ \bibinfo {pages} {177 -- 246}\BibitemShut {NoStop}%
\bibitem [{\citenamefont {Eberlein}\ \emph {et~al.}(2017)\citenamefont
  {Eberlein}, \citenamefont {Kasper}, \citenamefont {Sachdev},\ and\
  \citenamefont {Steinberg}}]{Eberlein2017Quantum}%
  \BibitemOpen
  \bibfield  {author} {\bibinfo {author} {\bibfnamefont {A.}~\bibnamefont
  {Eberlein}}, \bibinfo {author} {\bibfnamefont {V.}~\bibnamefont {Kasper}},
  \bibinfo {author} {\bibfnamefont {S.}~\bibnamefont {Sachdev}}, \ and\
  \bibinfo {author} {\bibfnamefont {J.}~\bibnamefont {Steinberg}},\ }\href
  {\doibase 10.1103/PhysRevB.96.205123} {\bibfield  {journal} {\bibinfo
  {journal} {Phys. Rev. B}\ }\textbf {\bibinfo {volume} {96}},\ \bibinfo
  {pages} {205123} (\bibinfo {year} {2017})}\BibitemShut {NoStop}%
\bibitem [{\citenamefont {{Bhattacharya}}\ \emph {et~al.}(2019)\citenamefont
  {{Bhattacharya}}, \citenamefont {{Jatkar}},\ and\ \citenamefont
  {{Sorokhaibam}}}]{Bhattacharya2019Quantum}%
  \BibitemOpen
  \bibfield  {author} {\bibinfo {author} {\bibfnamefont {R.}~\bibnamefont
  {{Bhattacharya}}}, \bibinfo {author} {\bibfnamefont {D.~P.}\ \bibnamefont
  {{Jatkar}}}, \ and\ \bibinfo {author} {\bibfnamefont {N.}~\bibnamefont
  {{Sorokhaibam}}},\ }\href {\doibase 10.1007/JHEP07(2019)066} {\bibfield
  {journal} {\bibinfo  {journal} {J. High Energy Phys.}\ }\textbf {\bibinfo
  {volume} {2019}},\ \bibinfo {pages} {66} (\bibinfo {year}
  {2019})}\BibitemShut {NoStop}%
\bibitem [{\citenamefont {Kuhlenkamp}\ and\ \citenamefont
  {Knap}(2020)}]{Kuhlenkamp2020Periodically}%
  \BibitemOpen
  \bibfield  {author} {\bibinfo {author} {\bibfnamefont {C.}~\bibnamefont
  {Kuhlenkamp}}\ and\ \bibinfo {author} {\bibfnamefont {M.}~\bibnamefont
  {Knap}},\ }\href {\doibase 10.1103/PhysRevLett.124.106401} {\bibfield
  {journal} {\bibinfo  {journal} {Phys. Rev. Lett.}\ }\textbf {\bibinfo
  {volume} {124}},\ \bibinfo {pages} {106401} (\bibinfo {year}
  {2020})}\BibitemShut {NoStop}%
\bibitem [{\citenamefont {Haldar}\ \emph {et~al.}(2020)\citenamefont {Haldar},
  \citenamefont {Haldar}, \citenamefont {Bera}, \citenamefont {Mandal},\ and\
  \citenamefont {Banerjee}}]{Haldar2020Quench}%
  \BibitemOpen
  \bibfield  {author} {\bibinfo {author} {\bibfnamefont {A.}~\bibnamefont
  {Haldar}}, \bibinfo {author} {\bibfnamefont {P.}~\bibnamefont {Haldar}},
  \bibinfo {author} {\bibfnamefont {S.}~\bibnamefont {Bera}}, \bibinfo {author}
  {\bibfnamefont {I.}~\bibnamefont {Mandal}}, \ and\ \bibinfo {author}
  {\bibfnamefont {S.}~\bibnamefont {Banerjee}},\ }\href {\doibase
  10.1103/PhysRevResearch.2.013307} {\bibfield  {journal} {\bibinfo  {journal}
  {Phys. Rev. Research}\ }\textbf {\bibinfo {volume} {2}},\ \bibinfo {pages}
  {013307} (\bibinfo {year} {2020})}\BibitemShut {NoStop}%
\bibitem [{\citenamefont {Song}\ \emph {et~al.}(2017)\citenamefont {Song},
  \citenamefont {Jian},\ and\ \citenamefont {Balents}}]{Son2017Strongly}%
  \BibitemOpen
  \bibfield  {author} {\bibinfo {author} {\bibfnamefont {X.-Y.}\ \bibnamefont
  {Song}}, \bibinfo {author} {\bibfnamefont {C.-M.}\ \bibnamefont {Jian}}, \
  and\ \bibinfo {author} {\bibfnamefont {L.}~\bibnamefont {Balents}},\ }\href
  {\doibase 10.1103/PhysRevLett.119.216601} {\bibfield  {journal} {\bibinfo
  {journal} {Phys. Rev. Lett.}\ }\textbf {\bibinfo {volume} {119}},\ \bibinfo
  {pages} {216601} (\bibinfo {year} {2017})}\BibitemShut {NoStop}%
\bibitem [{\citenamefont {Abramowitz}\ and\ \citenamefont
  {Stegun}(1964)}]{Abramowitz1964Handbook}%
  \BibitemOpen
  \bibfield  {author} {\bibinfo {author} {\bibfnamefont {M.}~\bibnamefont
  {Abramowitz}}\ and\ \bibinfo {author} {\bibfnamefont {I.}~\bibnamefont
  {Stegun}},\ }\href@noop {} {\emph {\bibinfo {title} {Handbook of Mathematical
  Functions, With Formulas, Graphs, and Mathematical Tables,}}}\ (\bibinfo
  {publisher} {Dover Publications, Inc.},\ \bibinfo {address} {New York},\
  \bibinfo {year} {1964})\BibitemShut {NoStop}%
\bibitem [{\citenamefont {Malitsky}(2020)}]{Malitsky2019Golden}%
  \BibitemOpen
  \bibfield  {author} {\bibinfo {author} {\bibfnamefont {Y.}~\bibnamefont
  {Malitsky}},\ }\href {\doibase 10.1007/s10107-019-01416-w} {\bibfield
  {journal} {\bibinfo  {journal} {Mathematical Programming}\ }\textbf {\bibinfo
  {volume} {184}},\ \bibinfo {pages} {383} (\bibinfo {year}
  {2020})}\BibitemShut {NoStop}%
\bibitem [{\citenamefont {Cheipesh}\ \emph {et~al.}(2019)\citenamefont
  {Cheipesh}, \citenamefont {Pavlov}, \citenamefont {Scopelliti}, \citenamefont
  {Tworzyd\l{}o},\ and\ \citenamefont {Gnezdilov}}]{Cheipesh2019Reentrant}%
  \BibitemOpen
  \bibfield  {author} {\bibinfo {author} {\bibfnamefont {Y.}~\bibnamefont
  {Cheipesh}}, \bibinfo {author} {\bibfnamefont {A.~I.}\ \bibnamefont
  {Pavlov}}, \bibinfo {author} {\bibfnamefont {V.}~\bibnamefont {Scopelliti}},
  \bibinfo {author} {\bibfnamefont {J.}~\bibnamefont {Tworzyd\l{}o}}, \ and\
  \bibinfo {author} {\bibfnamefont {N.~V.}\ \bibnamefont {Gnezdilov}},\ }\href
  {\doibase 10.1103/PhysRevB.100.220506} {\bibfield  {journal} {\bibinfo
  {journal} {Phys. Rev. B}\ }\textbf {\bibinfo {volume} {100}},\ \bibinfo
  {pages} {220506(R)} (\bibinfo {year} {2019})}\BibitemShut {NoStop}%
\bibitem [{Note1()}]{Note1}%
  \BibitemOpen
  \bibinfo {note} {In thermal equilibrium the fluctuation dissipation relation
  states \cite {Kamenev2005Course}: $G_K(\omega ) = 2 i \protect \tmspace
  +\thinmuskip {.1667em} \protect \mathrm {Im}G_R(\omega ) \protect \qopname
  \relax o{tanh}\protect \genfrac {}{}{}0{\omega -\mu }{2T}$}\BibitemShut
  {NoStop}%
\bibitem [{\citenamefont {Parcollet}\ \emph {et~al.}(1998)\citenamefont
  {Parcollet}, \citenamefont {Georges}, \citenamefont {Kotliar},\ and\
  \citenamefont {Sengupta}}]{Parcollet1998Overscreened}%
  \BibitemOpen
  \bibfield  {author} {\bibinfo {author} {\bibfnamefont {O.}~\bibnamefont
  {Parcollet}}, \bibinfo {author} {\bibfnamefont {A.}~\bibnamefont {Georges}},
  \bibinfo {author} {\bibfnamefont {G.}~\bibnamefont {Kotliar}}, \ and\
  \bibinfo {author} {\bibfnamefont {A.}~\bibnamefont {Sengupta}},\ }\href
  {\doibase 10.1103/PhysRevB.58.3794} {\bibfield  {journal} {\bibinfo
  {journal} {Phys. Rev. B}\ }\textbf {\bibinfo {volume} {58}},\ \bibinfo
  {pages} {3794} (\bibinfo {year} {1998})}\BibitemShut {NoStop}%
\bibitem [{\citenamefont {Lunkin}\ \emph {et~al.}(2018)\citenamefont {Lunkin},
  \citenamefont {Tikhonov},\ and\ \citenamefont
  {Feigel'man}}]{Lunkin2018Sachdev}%
  \BibitemOpen
  \bibfield  {author} {\bibinfo {author} {\bibfnamefont {A.~V.}\ \bibnamefont
  {Lunkin}}, \bibinfo {author} {\bibfnamefont {K.~S.}\ \bibnamefont
  {Tikhonov}}, \ and\ \bibinfo {author} {\bibfnamefont {M.~V.}\ \bibnamefont
  {Feigel'man}},\ }\href {\doibase 10.1103/PhysRevLett.121.236601} {\bibfield
  {journal} {\bibinfo  {journal} {Phys. Rev. Lett.}\ }\textbf {\bibinfo
  {volume} {121}},\ \bibinfo {pages} {236601} (\bibinfo {year}
  {2018})}\BibitemShut {NoStop}%
\bibitem [{\citenamefont {Can}\ and\ \citenamefont
  {Franz}(2019)}]{Can2019Solvable}%
  \BibitemOpen
  \bibfield  {author} {\bibinfo {author} {\bibfnamefont {O.}~\bibnamefont
  {Can}}\ and\ \bibinfo {author} {\bibfnamefont {M.}~\bibnamefont {Franz}},\
  }\href {\doibase 10.1103/PhysRevB.100.045124} {\bibfield  {journal} {\bibinfo
   {journal} {Phys. Rev. B}\ }\textbf {\bibinfo {volume} {100}},\ \bibinfo
  {pages} {045124} (\bibinfo {year} {2019})}\BibitemShut {NoStop}%
\bibitem [{\citenamefont {Chowdhury}\ \emph {et~al.}(2018)\citenamefont
  {Chowdhury}, \citenamefont {Werman}, \citenamefont {Berg},\ and\
  \citenamefont {Senthil}}]{Chowdhury2018Translationally}%
  \BibitemOpen
  \bibfield  {author} {\bibinfo {author} {\bibfnamefont {D.}~\bibnamefont
  {Chowdhury}}, \bibinfo {author} {\bibfnamefont {Y.}~\bibnamefont {Werman}},
  \bibinfo {author} {\bibfnamefont {E.}~\bibnamefont {Berg}}, \ and\ \bibinfo
  {author} {\bibfnamefont {T.}~\bibnamefont {Senthil}},\ }\href {\doibase
  10.1103/PhysRevX.8.031024} {\bibfield  {journal} {\bibinfo  {journal} {Phys.
  Rev. X}\ }\textbf {\bibinfo {volume} {8}},\ \bibinfo {pages} {031024}
  (\bibinfo {year} {2018})}\BibitemShut {NoStop}%
\bibitem [{\citenamefont {Tikhanovskaya}\ \emph {et~al.}(2021)\citenamefont
  {Tikhanovskaya}, \citenamefont {Guo}, \citenamefont {Sachdev},\ and\
  \citenamefont {Tarnopolsky}}]{Tikhanovskaya2020Excitation}%
  \BibitemOpen
  \bibfield  {author} {\bibinfo {author} {\bibfnamefont {M.}~\bibnamefont
  {Tikhanovskaya}}, \bibinfo {author} {\bibfnamefont {H.}~\bibnamefont {Guo}},
  \bibinfo {author} {\bibfnamefont {S.}~\bibnamefont {Sachdev}}, \ and\
  \bibinfo {author} {\bibfnamefont {G.}~\bibnamefont {Tarnopolsky}},\ }\href
  {\doibase 10.1103/PhysRevB.103.075141} {\bibfield  {journal} {\bibinfo
  {journal} {Phys. Rev. B}\ }\textbf {\bibinfo {volume} {103}},\ \bibinfo
  {pages} {075141} (\bibinfo {year} {2021})}\BibitemShut {NoStop}%
\bibitem [{\citenamefont {Levitov}\ and\ \citenamefont
  {Lesovik}(1993)}]{Levitov1993Charge}%
  \BibitemOpen
  \bibfield  {author} {\bibinfo {author} {\bibfnamefont {L.~S.}\ \bibnamefont
  {Levitov}}\ and\ \bibinfo {author} {\bibfnamefont {G.~B.}\ \bibnamefont
  {Lesovik}},\ }\href
  {http://www.jetpletters.ac.ru/ps/1186/article_17907.shtml} {\bibfield
  {journal} {\bibinfo  {journal} {JETP Lett.}\ }\textbf {\bibinfo {volume}
  {58}},\ \bibinfo {pages} {230} (\bibinfo {year} {1993})}\BibitemShut
  {NoStop}%
\bibitem [{\citenamefont {Levitov}\ \emph {et~al.}(1996)\citenamefont
  {Levitov}, \citenamefont {Lee},\ and\ \citenamefont
  {Lesovik}}]{Levitov1996Electron}%
  \BibitemOpen
  \bibfield  {author} {\bibinfo {author} {\bibfnamefont {L.~S.}\ \bibnamefont
  {Levitov}}, \bibinfo {author} {\bibfnamefont {H.}~\bibnamefont {Lee}}, \ and\
  \bibinfo {author} {\bibfnamefont {G.~B.}\ \bibnamefont {Lesovik}},\ }\href
  {\doibase 10.1063/1.531672} {\bibfield  {journal} {\bibinfo  {journal}
  {Journal of Mathematical Physics}\ }\textbf {\bibinfo {volume} {37}},\
  \bibinfo {pages} {4845} (\bibinfo {year} {1996})}\BibitemShut {NoStop}%
\bibitem [{Note2()}]{Note2}%
  \BibitemOpen
  \bibinfo {note} {In equilibrium the fluctuation dissipation relation holds
  $G_K(\omega )=-2\pi i (1-2 n_{SY\protect \tmspace -\thinmuskip
  {.1667em}K}(\omega )) \nu _{SY\protect \tmspace -\thinmuskip
  {.1667em}K}(\omega )$, $Q_K(\omega )=-2\pi i (1-2 n_{res}(\omega )) \nu
  _{res}(\omega )$, where $n_{SY\protect \tmspace -\thinmuskip {.1667em}K}$ and
  $n_{res}$ are the Fermi distribution functions and $\nu _{SY\protect \tmspace
  -\thinmuskip {.1667em}K}=-\protect \frac {1}{\pi } \protect \mathrm {Im} G_R$
  and $\nu _{res}=-\protect \frac {1}{\pi } \protect \mathrm {Im} Q_R$ are the
  densities of states. Substituting those to Eq. (\ref {It}), one gets a
  familiar Fermi golden rule formula for the tunneling current \cite
  {Kamenev2005Course}: \\ $ \protect \mathcal {I} = 2\pi \protect \sqrt
  {p}\lambda ^2 \protect \tmspace -\thinmuskip {.1667em} \DOTSI \intop
  \ilimits@ \protect \tmspace -\thinmuskip {.1667em} d\omega \protect \tmspace
  +\thinmuskip {.1667em} \nu _{SY\protect \tmspace -\thinmuskip
  {.1667em}K}(\omega ) \nu _{res}(\omega ) (n_{SY\protect \tmspace -\thinmuskip
  {.1667em}K}(\omega ) - n_{res}(\omega )) $.}\BibitemShut {Stop}%
\bibitem [{\citenamefont {Varma}\ \emph {et~al.}(2002)\citenamefont {Varma},
  \citenamefont {Nussinov},\ and\ \citenamefont {van
  Saarloos}}]{Varma2002Singular}%
  \BibitemOpen
  \bibfield  {author} {\bibinfo {author} {\bibfnamefont {C.}~\bibnamefont
  {Varma}}, \bibinfo {author} {\bibfnamefont {Z.}~\bibnamefont {Nussinov}}, \
  and\ \bibinfo {author} {\bibfnamefont {W.}~\bibnamefont {van Saarloos}},\
  }\href {\doibase 10.1016/s0370-1573(01)00060-6} {\bibfield  {journal}
  {\bibinfo  {journal} {Physics Reports}\ }\textbf {\bibinfo {volume} {361}},\
  \bibinfo {pages} {267} (\bibinfo {year} {2002})}\BibitemShut {NoStop}%
\end{thebibliography}%

\end{document}